\documentclass[trackchanges]{aastex701}

\begin{document}

\title{Revealing the Spatial Distribution of Dust and Plasma Solitary Waves in Jupiter’s Innermost Magnetosphere Using Juno Waves Observations}

\author[orcid=0009-0000-9676-1733]{Wenyu Zhang}
\affiliation{UCLA Department of Earth, Planetary, and Space Sciences, 595 Charles E Young Dr East, Los Angeles, CA 90095}
\email{wenyuzhang@ucla.edu}  

\author[orcid=0000-0002-6917-8363]{Hao Cao} 
\affiliation{UCLA Department of Earth, Planetary, and Space Sciences, 595 Charles E Young Dr East, Los Angeles, CA 90095}
\affiliation{UCLA SPACE Institute, 603 Charles E Young Dr East, Los Angeles, CA 90095}
\email{hcao@epss.ucla.edu}

\author[orcid=0000-0002-5471-6202]{William S. Kurth}
\affiliation{University of Iowa, Department of Physics and Astronomy}
\email{william-kurth@uiowa.edu}

\author[0000-0002-3093-458X]{Mitchell M. Shen}
\affiliation{Princeton University, Department of Astrophysical Sciences}
\email{mitchellshen@princeton.edu}

\author[0009-0000-9262-8775]{Darrelle Wilkinson}
\affiliation{University of Iowa, Department of Physics and Astronomy}
\email{darrelle-wilkinson@uiowa.edu}

\author[0000-0003-3183-4289]{Donglai Ma}
\affiliation{UCLA Department of Earth, Planetary, and Space Sciences, 595 Charles E Young Dr East, Los Angeles, CA 90095}
\email{dma96@epss.ucla.edu}

\author[0000-0003-0796-4251]{Martin Brennan}
\affiliation{Jet Propulsion Laboratory}
\email{Martin.Brennan@jpl.nasa.gov}

\author[0000-0001-9200-9878]{George Hospodarsky}
\affiliation{University of Iowa, Department of Physics and Astronomy}
\email{george-hospodarsky@uiowa.edu}

\author[0000-0002-9115-0789]{Scott J. Bolton}
\affiliation{Southwest Research Institute}
\email{scott.bolton@swri.org }



\begin{abstract}
Jupiter's faint ring system contains micron- and submicron-sized dust whose dynamics are influenced by the planet's rapidly rotating magnetic field. Optical observations have established the morphology of the main ring and halo, but tenuous dust close to Jupiter remains difficult to characterize. We analyze nearly a decade of in-situ Juno Waves electric-field measurements using a two-stage convolutional neural network to identify and classify dust impacts, solitary waves, and other plasma waves. These in-situ dust-impact detections reveal a radially evolving structure extending well inside the traditionally observed halo. In the innermost region, dust is concentrated in a compact, equatorially symmetric layer, indicating that the optical inner boundary does not mark the end of the dust population. Farther outward, the layer expands vertically into a broad halo before contracting toward the main ring, directly revealing the toroidal morphology of the Jovian halo. Dust number density is enhanced within the halo and near the main-ring--halo transition. The vertical distribution also develops hemispheric asymmetries that vary with radial distance, consistent with an influence of Jupiter's magnetic geometry on charged-dust dynamics. In contrast, solitary waves occur predominantly at high latitudes, and their mapped magnetic footprints show a strong spatial correspondence with Jovian auroral structures, particularly the main auroral oval. Together, these in-situ measurements extend the remote-sensing picture of Jupiter's rings and reveal previously unresolved dust and plasma structures in its innermost magnetosphere.
\end{abstract}




\section{Introduction}
\label{sec:introduction}

All four giant planets in the Solar System possess ring systems, but their
physical properties differ substantially. Saturn is surrounded by a massive,
optically thick ring system dominated by water-ice particles ranging from
centimeter to meter scales and exhibiting structure over a wide range of
spatial scales \citep{cuzzi_evolving_2010}. Uranus is encircled by a system
of narrow, dark rings first discovered through stellar occultations and
subsequently characterized in detail during the Voyager~2 encounter
\citep{elliot_rings_1977,smith_voyager_1986}. Neptune likewise possesses a
system of dark and dusty rings, including narrow rings and azimuthally
confined arcs that were resolved by Voyager~2
\citep{smith_voyager_1989}. Jupiter's rings, first directly imaged by
Voyager~1 in 1979, are much fainter and are characterized by extremely low
optical depths and a substantial population of micron-sized dust
\citep{smith_jupiter_1979,showalter_jupiters_1987}. The diversity of these
systems demonstrates that planetary rings span regimes ranging from dense,
collisionally dominated disks to highly tenuous dusty structures in which
nongravitational forces can strongly influence individual particle
trajectories \citep{burns_formation_1999,cuzzi_evolving_2010}.

Indirect evidence for a Jovian ring preceded its direct imaging discovery.
During the Pioneer~11 encounter, energetic-particle measurements showed
pronounced charged-particle ``dropouts'' in Jupiter's inner magnetosphere,
some of which were interpreted as possible absorption by previously
undetected ring material \citep{fillius_radiation_1975}. Voyager~1
subsequently provided the first direct image of a particulate ring around
Jupiter in March 1979 \citep{smith_jupiter_1979}, and Voyager~2 observations
soon confirmed and expanded the initial picture of the ring system
\citep{owen_jupiters_1979}. Reanalysis of the Voyager images established
that the Jovian ring system contains a relatively bright and geometrically
thin main ring together with a much more vertically extended inner halo
\citep{jewitt_jovian_1981,showalter_jupiters_1987}. A still fainter outer component, later named
the gossamer ring, was identified in enhanced Voyager images by
\citet{showalter_discovery_1985}.

The main ring of Jupiter extends from approximately $1.72$ to
$1.81~R_J$, with an abrupt outer boundary near the orbit of Adrastea and
Metis embedded within its outer portion
\citep{showalter_jupiters_1987,burns_jupiters_2004}. Despite its radial
width of approximately 7000~km, the bright main ring is geometrically very
thin, with a north--south thickness of no more than $\sim30$~km in
backscattered light \citep{showalter_jupiters_1987,burns_formation_1999}.
Its normal optical depth is only a few $\times10^{-6}$
\citep{showalter_jupiters_1987,burns_jupiters_2004}. The main ring contains
both dust-sized particles and a population of substantially larger parent
bodies, with its appearance changing strongly with viewing phase angle
because forward scattering emphasizes the small-grain component
\citep{showalter_jupiters_1987,brooks_size_2004}. New Horizons later
detected localized clumps within the main ring but no additional
kilometer-scale satellites, reinforcing the view that the observed dust
must be continually replenished from a population of small embedded source
bodies and the known inner moons \citep{showalter_clump_2007}.

Immediately interior to the main ring lies the Jovian halo. Voyager images
showed that the ring material expands rapidly away from the equatorial plane
toward smaller radial distances, reaching a full vertical thickness of order
$2\times10^{4}$~km ($0.28R_J$) before becoming too faint to detect in the innermost
region \citep{showalter_jupiters_1987}. The halo has an optical depth of order
$10^{-6}$ and is commonly described as occupying approximately
$1.4$--$1.72~R_J$, although increasingly faint material may extend farther
inward than this nominal optical boundary
\citep{burns_jupiters_2004,ye_juno_2020}. Galileo imaging subsequently
resolved the transition from the thin main ring into the vertically extended
halo and confirmed that the ring brightness is dominated in forward
scattering by small dust grains
\citep{ockert-bell_structure_1999}. Ground-based Keck observations and later
spacecraft measurements further refined the radial and particle-size
structure of the main-ring--halo system
\citep{de_pater_keck_2008}.

Exterior to the main ring, the gossamer ring system extends to approximately
$3.1$--$3.2~R_J$. Galileo observations revealed that what initially
appeared to be a single diffuse outer ring consists primarily of
two overlapping components associated with the orbits of Amalthea and Thebe
\citep{ockert-bell_structure_1999}. The radial and vertical morphology of
these rings is broadly consistent with dust released from the surfaces of
the corresponding moons and subsequently transported inward
\citep{burns_formation_1999,showalter_properties_2008}.

The short dynamical lifetime of micron-sized grains implies that Jupiter's
dusty rings cannot be primordial static structures and must be continuously
replenished. The commonly adopted source mechanism is hypervelocity
micrometeoroid bombardment of the small inner satellites and larger parent
bodies embedded within the rings, which ejects dust above the local escape
velocity \citep{burns_formation_1999,burns_jupiters_2004}. For the gossamer
rings, the correspondence between the ring boundaries and the Amalthea and
Thebe orbits provides particularly strong evidence for this ejecta-source
model \citep{burns_formation_1999,showalter_properties_2008}. Similar
collisional production is thought to supply the dust component of the main
ring, with Metis, Adrastea, and smaller unresolved parent bodies serving as
potential sources \citep{burns_formation_1999,showalter_clump_2007}.

Once produced, micron- and submicron-sized grains in the Jovian ring
system become electrically charged through photoelectric emission and
interactions with the surrounding plasma
\citep{horanyi_structure_1996,horanyi_plasma_2010}. In the main-ring--halo
region, \citet{horanyi_plasma_2010} argued that superthermal photoelectrons
escaping from Jupiter's ionosphere are likely to provide the dominant ambient
electron source, and their model predicts positively charged grains with
equilibrium surface potentials increasing from approximately 6.5~V at
$1.4~R_J$ to about 8~V at $1.8~R_J$. Once charged, the grains are strongly
affected by Jupiter's rapidly rotating, non-axisymmetric magnetic field, and
their dynamics can be modified by Lorentz resonances
\citep{burns_lorentz_1985,burns_jupiters_2004}. Several prominent ring
boundaries lie close to first-order Lorentz resonances: the 2:1 resonance near
$1.4~R_J$ lies close to the inner optical boundary of the halo, the 3:2
resonance near $1.71~R_J$ corresponds closely to the transition between the
vertically extended halo and the geometrically thin main ring, and the 4:3
resonance occurs near $1.85~R_J$, just exterior to the bright main ring
\citep{burns_lorentz_1985,ockert-bell_structure_1999,
de_pater_keck_2008,horanyi_plasma_2010}. Because the electromagnetic
response increases with grain charge-to-mass ratio, smaller grains can
undergo substantially larger radial and vertical excursions, providing a
natural mechanism for the size-dependent structure of Jupiter's main ring
and halo \citep{burns_lorentz_1985,horanyi_plasma_2010}.

Most previous observations of Jupiter's main ring and halo have relied on
optical and near-infrared remote sensing
\citep{showalter_jupiters_1987,ockert-bell_structure_1999,
de_pater_keck_2008}. These measurements established the global radial and
vertical morphology of the ring system and provided important constraints on
its particle population. Juno, however, offers a complementary in-situ view.
The spacecraft carries the Waves investigation, whose electric-field antenna
can detect voltage pulses produced when dust grains impact the spacecraft at
hypervelocity \citep{kurth_juno_2017}. Such impacts vaporize and ionize the
projectile and part of the spacecraft surface, producing an expanding plasma
cloud that perturbs the spacecraft and antenna potentials
\citep{gurnett_micron-sized_1983,meyer-vernet_detecting_2001}. In contrast
to optical measurements, which infer dust properties from scattered light
integrated along the line of sight, this technique directly samples the local
dust population and can remain sensitive to sparse and very small particles,
particularly at high impact velocities.

The use of radio and plasma-wave electric antennas as dust detectors dates
back to the Voyager~2 encounter with Saturn, when intense impulsive signals
near the ring plane were attributed to impacts of micron-sized particles
\citep{gurnett_micron-sized_1983}. The technique was subsequently developed
with Cassini RPWS, where dust-impact voltage pulses were used to measure
impact rates and spatial dust distributions in Saturn's magnetosphere
\citep{kurth_cassini_2006,ye_properties_2014,ye_-situ_2016,ye_understanding_2019}. Similar measurements have since been made in interplanetary and planetary environments with
STEREO Waves \citep{meyer-vernet_dust_2009}, Solar Orbiter RPW \citep{zaslavsky_interplanetary_2012,kvammen_machine_2023}, and MAVEN LPW \citep{andersson_dust_2015, ijaz_study_2024}. These studies have established electric-field
instruments as effective in-situ dust detectors over a wide range of plasma
environments and particle sizes.

At Jupiter, \citet{ye_juno_2020} first applied this technique to Juno Waves
observations from the initial 21 perijoves and identified dust-impact signals
during 20 ring-plane crossings. During these early orbits, Juno crossed the
equatorial plane at approximately $1.04$--$1.1~R_J$, well interior to the
$\sim1.4~R_J$ inner boundary at which the halo becomes undetectable in
optical observations. The detected impacts were strongly concentrated near
the equatorial plane, with characteristic half-widths of approximately
2000--3000~km, and were interpreted as micron-scale dust with number
densities of order $10^{-6}~{\rm m^{-3}}$
\citep{ye_juno_2020}. These measurements provided the first in-situ evidence
for a dust population deeply inward of Jupiter's optically observed ring system
and demonstrated the unique capability of Juno Waves to probe the innermost
extension of the Jovian halo.

The high-time-resolution electric-field measurements used for dust detection
also contain a variety of physically interesting transient plasma signals.
Some of these signals can exhibit short-duration impulsive or oscillatory
structures that partially resemble dust-impact waveforms, making them
important competing populations in dust identification. Among the most
prominent examples are electrostatic solitary waves, which appear as
large-amplitude, localized electric-field structures and have been observed
in a variety of planetary plasma environments. Such structures have been
reported in Earth's auroral regions, where they are associated with
field-aligned electron beams and downward field-aligned currents
\citep{ergun_fast_1998}, and at Saturn using Cassini electric-field
measurements \citep{williams_electrostatic_2006}. At Jupiter, Juno Waves
burst observations have revealed large-amplitude solitary waves in the
high-latitude auroral region, particularly in Zone-II, which is associated
with downward field-aligned currents and broadband electron populations
\citep{sulaiman_jupiters_2022}. The Juno Waves measurements also contain other short-duration plasma wave emissions. Observations in Jupiter's auroral regions and along magnetic flux
tubes connected to Io have identified ion-cyclotron, Alfv\'en, and
whistler-mode waves, demonstrating that several physically distinct wave
modes can occur within the same low-frequency electric-field data
\citep{kurth_whistler_2018,sulaiman_wave-particle_2020,
sulaiman_jupiters_2022}. These plasma-wave signals are scientifically
valuable in their own right, but their presence also complicates automated
dust detection because transient wave packets can sometimes exhibit
time-domain morphologies similar to impulsive dust signatures. This
motivates treating the Juno Waves burst data as a multi-class transient-signal
population rather than simply separating dust impacts from an undifferentiated
background.

In this work, we use nearly a decade of Juno Waves LFR\_LO\_E burst
observations spanning PJ01--PJ83 to investigate transient electric-field
signals in Jupiter's inner magnetosphere. We develop a two-stage
convolutional neural network (CNN) framework that first identifies transient
waveforms and subsequently classifies them into three populations:
\textit{Dust Impact}, \textit{Solitary Wave}, and \textit{Other Waves}.
The extended Juno data set enables us to map the dust impact rate and inferred
number density throughout Jupiter's innermost magnetosphere, reveal its
radial and vertical structure well inside the optically observed halo, and
examine the size-dependent organization of the detected grains. The
multi-class classification also preserves the nondust transient populations,
allowing their spatial distributions---particularly the high-latitude
solitary-wave population---to be investigated simultaneously.

\section{Data and Methods}

\subsection{Juno Waves Burst Observations}
\label{sec:juno_waves_data}

NASA's Juno spacecraft entered polar orbit around Jupiter in July 2016
\citep{bolton_juno_2017}. Juno follows a highly eccentric orbit that
repeatedly carries the spacecraft from the outer Jovian magnetosphere to
close perijove passages through Jupiter's upper ionosphere
\citep{kurth_electron_2025}. These perijove (PJ) trajectories provide a
particularly valuable observing geometry for the present study: Juno
repeatedly samples the innermost magnetosphere at very small radial distances
while traversing a wide range of latitudes. The multi-year sequence of PJ
orbits therefore enables in-situ investigation of the spatial distributions
of dust and transient plasma-wave phenomena in regions that are difficult to
sample with more equatorially confined spacecraft trajectories.

Because of Juno's polar orbit, the spacecraft crosses Jupiter's equatorial
plane twice during each orbit, once on the outbound descending portion and
again on the inbound ascending portion. As Juno's orbit precessed over the
mission, the perijove location gradually shifted toward higher northern
latitudes. Correspondingly, the descending equatorial crossing moved
progressively farther from Jupiter, while the ascending crossing moved
progressively closer to the planet. For the equatorial halo observations considered in this study, we use the
descending crossings. Across the PJ intervals analyzed here, these crossings
progressively extend outward from the innermost dust environment and span
radial distances of approximately $1.05$--$1.636~R_J$. This orbital
evolution allows the Juno Waves observations to progressively sample the
dust population from the inner equatorial region toward the Jovian halo and
main-ring region. The corresponding ascending crossings are not included
in the present equatorial halo-crossing analysis.

The Juno Waves investigation measures electric and magnetic fluctuations over
a broad frequency range \citep{kurth_juno_2017}. Electric-field measurements
are obtained using a V-shaped dipole antenna composed of two 2.728-m elements.
In this study, we use calibrated burst waveforms from the Low Frequency
Receiver Lo-E channel (LFR\_LO\_E), which measures electric-field fluctuations
from approximately 50~Hz to 20~kHz. A standard LFR\_LO\_E burst record
contains 6144 samples acquired at 50~kHz, corresponding to a waveform duration
of 0.12288~s. Unlike the regular survey products, these burst measurements
preserve the high-time-resolution electric-field waveform and are therefore
well suited for identifying impulsive dust-impact signatures and other
short-duration transient structures. We use the calibrated full-resolution
Juno Waves burst data
(\texttt{JNO-E/J/SS-WAV-3-CDR-BSTFULL-V2.0}) \citep{kurth_juno_2024} distributed through NASA's
Planetary Data System, with each record providing electric-field amplitude as
a function of time for input to the CNN classification described in
Section~\ref{sec:cnn_classification}. The assembled LFR\_LO\_E catalog contains 961,169 burst waveform records spanning PJ01--PJ83, from 27 August 2016 to 2 May 2026. A total of 81 perijove intervals are represented; PJ02 and PJ76 are absent from the assembled data set.

\subsection{CNN Classification and Performance}
\label{sec:cnn_classification}

\begin{figure*}[htb!]
    \centering
    \includegraphics[width=0.95\textwidth]{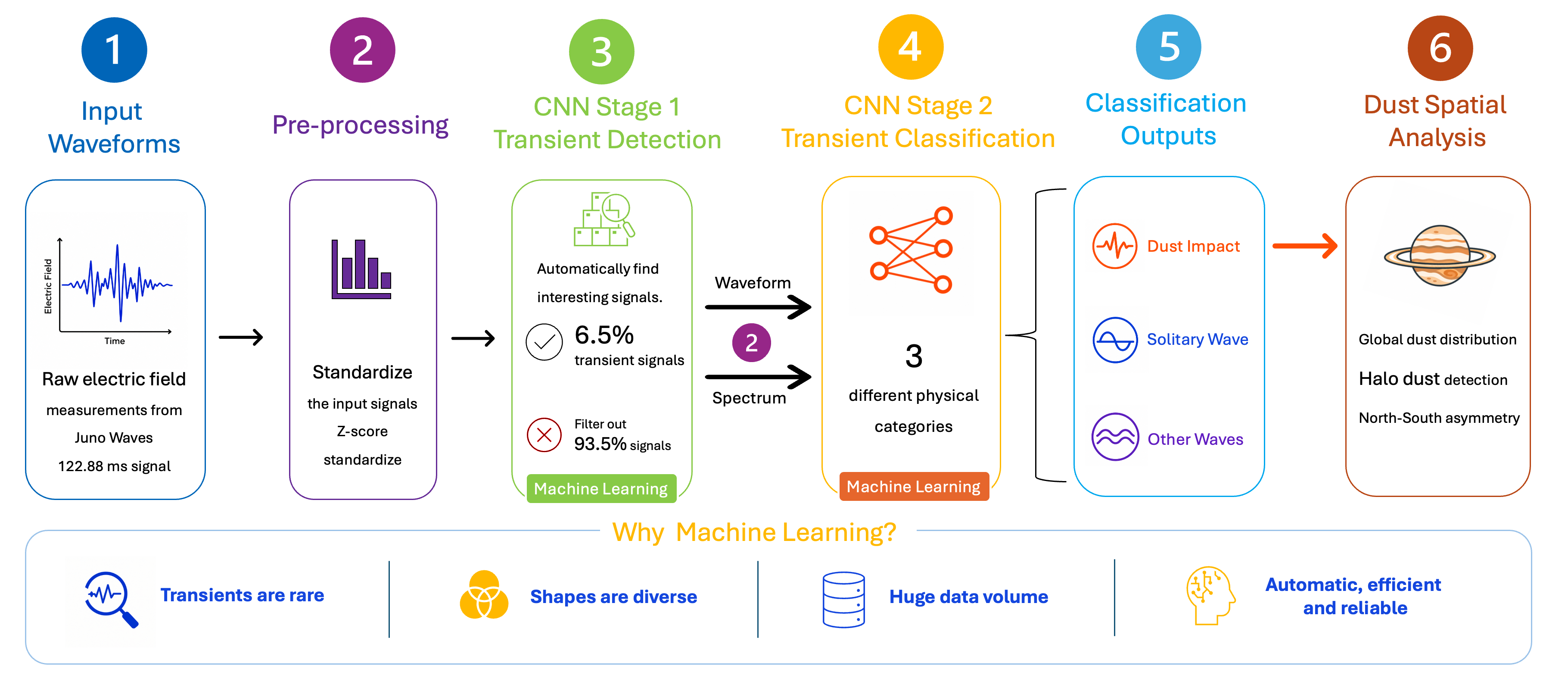}
    \caption{
    Schematic overview of the two-stage CNN classification framework.
    Each Juno Waves LFR\_LO\_E waveform is first standardized independently.
    Stage~1 identifies transient signals from the full waveform data set.
    The retained candidates are then represented by both their time-domain
    waveforms and Fourier amplitude spectra and passed to Stage~2, which
    classifies each signal as \textit{Dust Impact}, \textit{Solitary Wave},
    or \textit{Other Waves}. The resulting populations are used in the
    subsequent spatial analyses.
    }
    \label{fig:cnn_pipeline}
\end{figure*}

The large volume of Juno Waves burst observations makes manual identification
of transient electric-field signals impractical. Convolutional neural
networks (CNNs) are well suited for waveform-classification problems because
they can learn characteristic local structures directly from time-series
data while requiring relatively few trainable parameters
\citep{lecun_gradient-based_1998}. Their applicability to electric-field
dust detection has also been demonstrated using Solar Orbiter/RPW
observations, where CNN-based classification was used to distinguish
dust-impact signatures from other transient waveform populations
\citep{kvammen_machine_2023}. Hence, we developed a two-stage
one-dimensional CNN framework for the Juno Waves LFR\_LO\_E burst data.
The overall workflow is summarized in Figure~\ref{fig:cnn_pipeline}.

Prior to CNN classification, each waveform was standardized independently
using a generalized record-wise Z-score transformation,
\begin{equation}
E_i^{*}
=
\frac{E_i-C(E)}
{\sigma_E+\epsilon},
\label{eq:waveform_standardization}
\end{equation}
where $C(E)$ represents the central value of the individual waveform and
$\sigma_E$ is its standard deviation. For Stage~1,
$C(E)=\overline{E}$ and $\epsilon=0$, corresponding to conventional
Z-score standardization. For the Stage~2 time-domain input,
$C(E)=\mathrm{median}(E)$ and $\epsilon=10^{-6}$, which reduces the
sensitivity of the estimated waveform baseline to large impulsive
excursions. The Stage~2 Fourier amplitude spectrum was standardized
separately using the same formulation with mean centering and
$\epsilon=10^{-6}$.

Stage~1 serves as a transient-signal detector. The standardized
6144-sample waveform is supplied to a compact 1D CNN containing three
successive convolutional blocks followed by global average pooling and a
fully connected output layer. The model contains 44,737 parameters and was
trained on 10,865 labeled waveforms, including 10,372 background waveforms
and 493 transient signals. Balanced class weights were used to account for
the strong class imbalance. Training used the Adam optimizer
\citep{kingma_adam_2017} with binary cross-entropy loss. The final output
layer uses a sigmoid activation function and returns a transient-signal
probability between 0 and 1. Waveforms with probabilities greater than
0.6443 were retained as transient candidates for Stage~2.

Stage~2 performs the multi-class classification of the retained transient
signals. To capture complementary waveform characteristics, the classifier
uses both the standardized time-domain waveform and its Fourier amplitude
spectrum. The spectrum is calculated after mean removal and Hann windowing
and is restricted to 50~Hz--20~kHz. The waveform and spectral inputs are
processed by separate 1D CNN branches, and their extracted features are
combined before the final classification layer. The complete Stage~2 model
contains 67,267 parameters. Its three-node output layer uses a softmax
activation function to assign each transient signal to one of three classes:
\textit{Dust Impact}, \textit{Solitary Wave}, or \textit{Other Waves}.

Because the labeled transient data set is relatively small and imbalanced
among the three classes, Stage~2 was trained using the Adam optimizer
\citep{kingma_adam_2017} and a multi-class focal loss
\citep{lin_focal_2017},
\begin{equation}
\mathcal{L}
=
-\sum_{c=1}^{3}
\alpha_c y_c(1-p_c)^{\gamma}\log p_c ,
\label{eq:focal_loss}
\end{equation}
where $\gamma=2$ and
$\alpha=(1,3,2)$ for \textit{Dust Impact},
\textit{Solitary Wave}, and \textit{Other Waves}, respectively. The
effective labeled data set contained 669 matched transient examples and was
divided using a stratified 60\%/20\%/20\% split into training, validation,
and independent test subsets. The final classifier was selected according
to its validation-set performance without using the independent test set for
model selection.

The Stage~1 operating threshold of 0.6443 yields a precision of 0.9004, a
recall of 0.9533, and an average precision of 0.9467. Because these values
were evaluated on the labeled data set used for threshold selection, they are
used primarily to characterize the screening behavior of Stage~1 rather
than as an independent estimate of model generalization. Stage~2 was
evaluated separately on an independent test set containing 134 transient
waveforms. The deployed classifier achieves an overall test accuracy of
0.9254 and a macro-averaged F1 score of 0.8576. The corresponding confusion
matrix is shown in Table~\ref{tab:stage2_confusion}.

\begin{table}[t]
\centering
\caption{
Confusion matrix of the Stage~2 multi-class CNN evaluated on the independent
test set. Bold diagonal entries represent correct classifications, while
off-diagonal entries represent misclassifications.
}
\label{tab:stage2_confusion}

\begin{tabular}{lccc}
\hline

\raisebox{-0.8\height}{\textbf{\textit{True Class}}}
& \multicolumn{3}{c}{\textbf{Classified Class}} \\

& \textbf{Dust Impact}
& \textbf{Solitary Wave}
& \textbf{Other Waves} \\

\hline

\textit{Dust Impact}
& \textbf{100} & 1 & 1 \\

\textit{Solitary Wave}
& 4 & \textbf{15} & 1 \\

\textit{Other Waves}
& 3 & 0 & \textbf{9} \\

\hline
\end{tabular}

\end{table}

The two-stage architecture is a key feature of our classification strategy.
Rather than requiring a single network to distinguish several rare transient
classes directly from the much larger background population, Stage~1 first
screens the full Juno Waves data set and retains waveforms containing
potentially interesting transient structure. Stage~2 can then focus on the
more subtle differences among physically distinct transient populations.
This hierarchical design substantially reduces the class-imbalance problem
faced by the multi-class classifier and limits the complexity required of
each individual model, thereby reducing the risk of overfitting when only a
modest number of manually labeled examples are available. Both CNNs are
intentionally compact, containing fewer than $7\times10^{4}$ parameters
each, yet the combined framework provides high-recall transient screening
and accurate multi-class classification. Importantly, this approach also
preserves solitary waves and other plasma-wave signals as scientifically
useful populations rather than treating all nondust transients simply as
background.

\section{Result}

\subsection{Global Distribution and Structure of Dust}
\label{sec:dust_global_distribution}

We first examine the spatial distribution of the CNN-identified dust population
using the impact rate,
\begin{equation}
R_{\rm impact}
=
\frac{N_{\rm impact}}{T_{\rm obs}},
\label{eq:impact_rate}
\end{equation}
where $N_{\rm impact}$ is the number of identified dust impacts and
$T_{\rm obs}$ is the accumulated Juno Waves burst observation time within
each spatial bin. This normalization accounts for the strongly nonuniform
spatial sampling produced by Juno's polar orbit.

\begin{figure*}[htb!]
    \centering
    \includegraphics[width=0.49\textwidth]
    {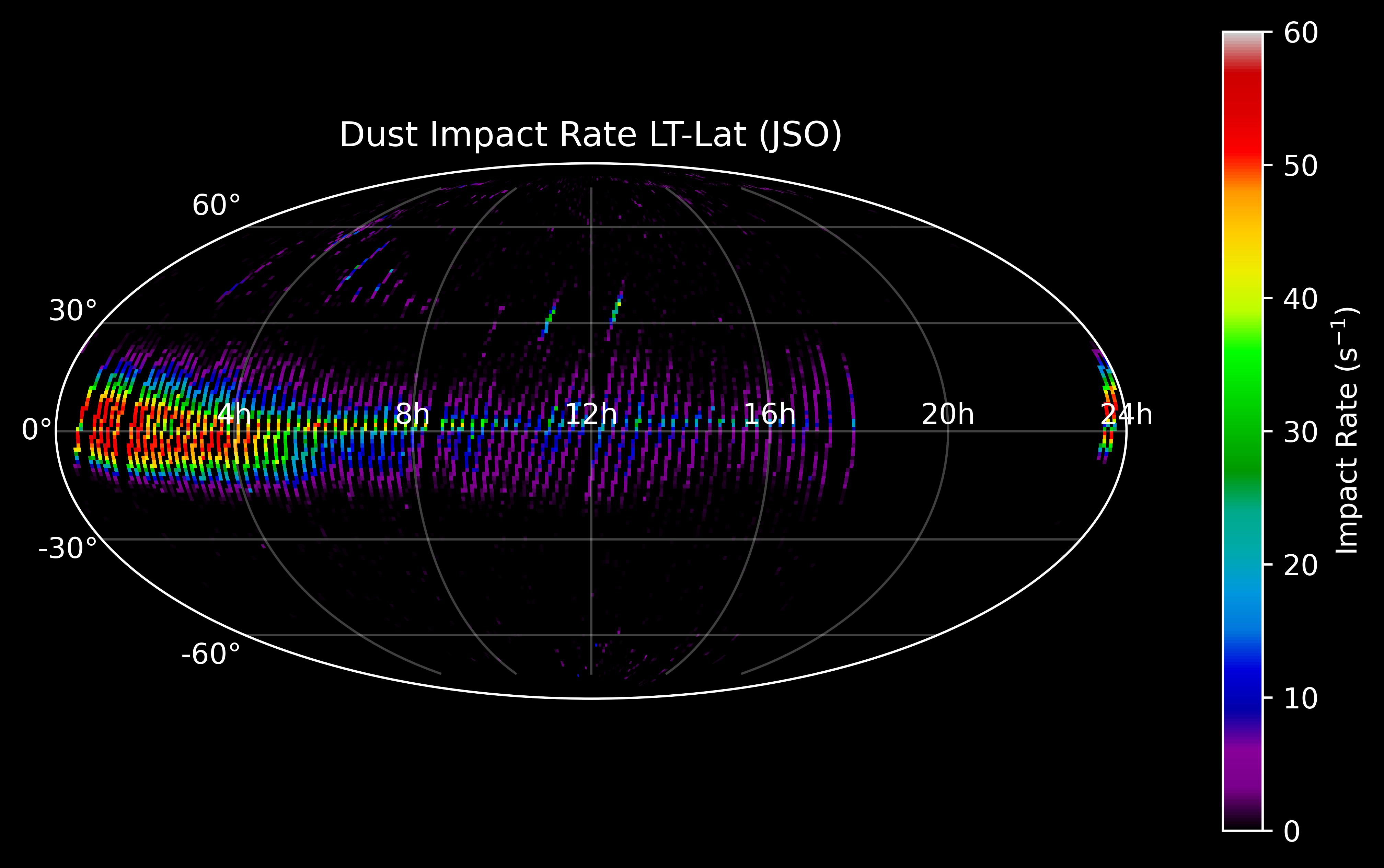}
    \hfill
    \includegraphics[width=0.49\textwidth]
    {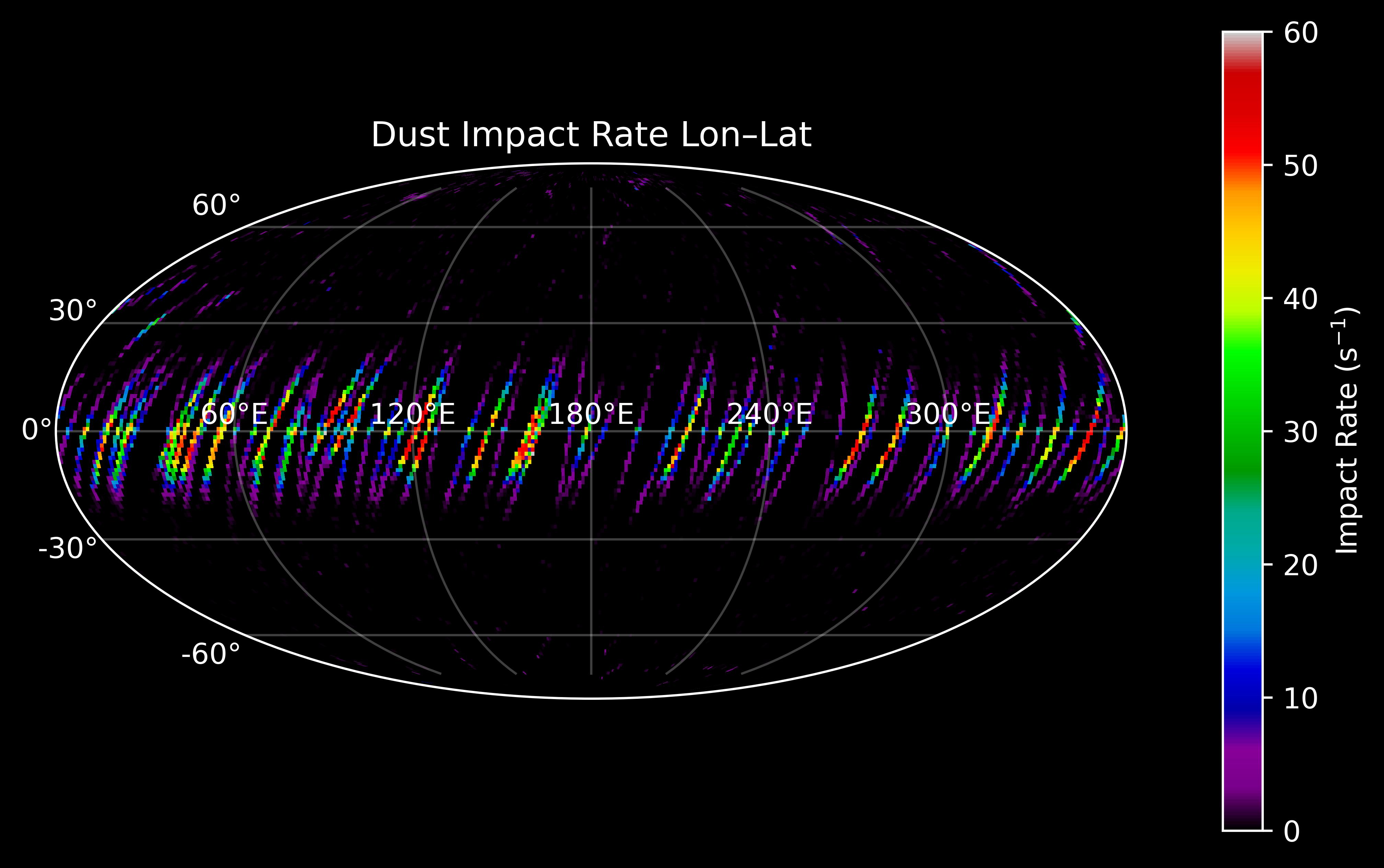}
    \caption{
    Global distribution of the dust impact rate identified from Juno Waves
    burst observations.
    Left: JSO local-time--latitude distribution.
    Right: System~III longitude--latitude distribution.
    The maps show that Juno broadly sampled the dust population over many
    longitudes and local-time sectors, while the highest impact rates remain
    strongly concentrated near Jupiter's equatorial plane.
    }
    \label{fig:dust_rate_global}
\end{figure*}

Figure~\ref{fig:dust_rate_global} shows the global impact-rate distribution
in System~III longitude--latitude and Jupiter--Sun--Orbit (JSO)
local-time--latitude coordinates. System~III is a Jupiter-fixed coordinate
system that rotates with the planet, whereas JSO is a Sun-referenced system
commonly used to describe the spacecraft geometry relative to the Sun and
Jupiter's orbital plane \citep{wang_jupiters_2023}. In System~III
coordinates, dust impacts are detected over nearly the full sampled
longitude range, indicating that the dust population is broadly distributed
around the planet. At the same time, both the System~III and JSO maps show a
pronounced concentration of high impact rates near the equatorial plane,
with local values reaching $\sim60~{\rm s^{-1}}$.

The JSO distribution also reflects the evolution of Juno's sampling geometry
during the mission. As Juno's orbit precessed, the perijove trajectories
progressively sampled different local-time sectors, evolving from the dusk
side toward the dayside and subsequently toward the dawn and nightside
sectors. Later perijoves also sampled dust at progressively larger radial
distances, including regions closer to the Jovian halo and main ring. Because
the vertical extent of the dust population increases with radial distance
over much of this region, the JSO map correspondingly shows detections over
a broader range of latitudes. Despite this geometrical evolution, the
equatorial plane remains the persistent location of the strongest dust
impact enhancement.

To further quantify the dust spatial distribution, we convert the detected
impacts into an inferred dust number density,
\begin{equation}
n_{\rm d}
=
\frac{N_{\rm impact}}
{\displaystyle\sum_i A_i v_i \Delta t},
\label{eq:dust_density}
\end{equation}
where $A_i$ and $v_i$ are the effective spacecraft impact area and relative
dust--spacecraft velocity associated with the $i$th observation,
respectively, and $\Delta t=0.12288$~s is the duration of an
LFR\_LO\_E burst waveform. To estimate the relative velocity, we assume that the dust grains move on circular Keplerian orbits around Jupiter in the prograde direction. The relative dust–spacecraft velocity is then calculated from the difference between the local Keplerian dust velocity and the Juno spacecraft velocity. The effective spacecraft impact area and
spacecraft velocity are described in Appendix~\ref{app:impact_area_speed}.

Figure~\ref{fig:dust_spatial_profiles} summarizes the large-scale spatial
and radial structure of the dust population. Panels (a) and (b) show the
impact rate and inferred number density in $r$--latitude coordinates,
respectively. Panels (c) and (d) show the number density distribution in
cylindrical $\rho$--$z$ coordinates as a two-dimensional density map and
as density contours, while panel (e) compares the cylindrical radial
profiles of the mid-plane, northern, and southern populations.

Inside approximately $1.2~R_J$, the dust population is strongly confined
to the equatorial plane and contains two localized high-density regions near
$\rho\sim1.10~R_J$ and $\rho\sim1.15~R_J$. Within these equatorial
enhancements, the impact rate reaches approximately $50~{\rm s^{-1}}$,
while the inferred number density is typically
$\sim1.5$--$2.0\times10^{-5}~{\rm m^{-3}}$. Both quantities decrease
rapidly away from the equatorial plane, indicating that the inner dust
population is vertically compact. The characteristic vertical HWHM in this
region is only $\sim0.03$--$0.08~R_J$, corresponding to approximately
$2{,}100$--$5{,}700$~km. These localized enhancements coincide with relatively small effective spacecraft impact areas and high spacecraft velocities (Appendix~\ref{app:impact_area_speed}), which may introduce additional uncertainty in the inferred number density. The enhancements should therefore be interpreted with some caution.

Beyond $\sim1.2~R_J$, the dust layer exhibits a clear
expansion--contraction pattern rather than a monotonic increase in vertical
extent. Moving outward from the inner equatorial population, the distribution
broadens rapidly and reaches its maximum observed vertical extent near
$\sim1.4~R_J$. In this region, the vertical HWHM increases to approximately
$0.367~R_J$ ($\sim2.62\times10^4$~km), corresponding to a full width at
half maximum of approximately $0.734~R_J$
($\sim5.25\times10^4$~km). This extent is equivalent to roughly
$\pm15^\circ$ in latitude at these radial distances. Beyond
$\sim1.4~R_J$, the dust layer becomes progressively narrower again toward
the main-ring region. The resulting cross section therefore has a broad,
toroidal morphology.

This large vertical extent is consistent with the broad dimensions of the
Jovian halo inferred from previous remote-sensing observations. Voyager
images established that the halo forms a vertically extended, toroidal dust
structure interior to the main ring \citep{showalter_jupiters_1987}.
Using Galileo imaging, \citet{ockert-bell_structure_1999} placed the inner
edge of the halo near $92{,}000\pm1{,}000$~km ($\sim1.29~R_J$) and found
that its diffuse upper and lower boundaries extend to approximately
$25{,}000$~km above and below the ring plane. Although the brighter central
portion of the halo has a substantially smaller characteristic FWHM, these
outer boundaries correspond to a total vertical extent of order
$5\times10^4$~km. Similar characteristic dimensions of order
$10^4$--$10^5$~km have been summarized in subsequent studies of the
Jovian halo \citep{burns_formation_1999,pater_rings_2018}. The maximum
vertical extent measured in situ by Juno Waves is therefore comparable to
the diffuse outer boundary of the halo inferred from imaging. Because
optical brightness and in situ number density represent different
observables, this agreement should be interpreted in terms of overall
spatial scale rather than as a direct comparison of profile widths.

The cylindrical $\rho$--$z$ distributions in
Figures~\ref{fig:dust_spatial_profiles}c and
\ref{fig:dust_spatial_profiles}d provide a complementary view of this
vertical evolution. The density map shows that the highest inferred number
densities remain concentrated close to the equatorial plane in the
innermost region, including the localized enhancements near
$\rho\sim1.10~R_J$ and $\rho\sim1.15~R_J$. Farther outward, the
distribution expands substantially in $z$. The corresponding isodensity
contours emphasize the changing cross-sectional morphology of the dust
population, from a compact inner equatorial layer to a much broader halo at
intermediate cylindrical distances and a subsequent contraction toward the
main-ring region.

To quantify the radial organization of the dust population in ring-related
geometry, we compute cylindrical radial profiles in bins of width
$\Delta\rho=0.01~R_J$. Figure~\ref{fig:dust_spatial_profiles}e compares
the northern ($0<z\leq0.5~R_J$), southern
($-0.5~R_J\leq z<0$), and near-midplane ($|z|\leq5{,}000$~km)
number density profiles.

The profiles reveal three characteristic radial regions. The first is a compact inner population at approximately $\rho=1.1$--$1.2~R_J$,
corresponding to the equatorially confined dust observed inward of the optical halo. A second, broader enhancement extends across approximately $\rho=1.3$--$1.5~R_J$, coincident with the vertically expanded halo identified in the $\rho$--$z$ distribution. Finally, the profiles show another enhancement near $\rho\sim1.7~R_J$, at the radial location of the main-ring--halo transition. Together, these features indicate that the radial number density distribution is organized into distinct inner,
halo, and main-ring-related regimes rather than varying monotonically with
cylindrical distance.

The hemispheric profiles further show that this radial organization is
accompanied by changes in vertical structure. Between approximately
$\rho=1.1$ and $1.2~R_J$, the northern and southern profiles are broadly
comparable, consistent with the approximately symmetric inner dust layer.
Farther outward, the northern density exceeds the southern density
throughout much of the $\rho\sim1.2$--$1.6~R_J$ interval, while the
near-midplane profile also remains elevated through the halo region. The
$2{:}1$ Lorentz resonance at $1.40~R_J$ lies within this broad halo
enhancement rather than coinciding with a distinct local maximum or break.

Near $\rho\sim1.7~R_J$, the southern profile shows a pronounced
enhancement, while the near-midplane profile also rises toward the
main-ring--halo transition and the $3{:}2$ Lorentz resonance at
$\sim1.71~R_J$. The sampling in this region, however, is strongly
hemispherically biased, with substantially more observations obtained
south of the equatorial plane. The pronounced southern enhancement should
therefore be interpreted with caution and may partly reflect the available
Juno sampling geometry. Moreover, even the near-midplane interval used here
extends to $|z|=5{,}000$~km, substantially larger than the vertical
thickness of the main ring, which is only tens to a few hundred kilometers
depending on the particle population
\citep{showalter_jupiters_1987,ockert-bell_structure_1999,pater_rings_2018}.
The outer enhancement therefore identifies a dust population at the radial
location of the main ring but should not be interpreted as a measurement of
the intrinsic main-ring midplane number density.

A pronounced minimum near $\rho\sim1.6~R_J$ coincides with changes in the
Waves attenuation configuration and is therefore not interpreted as a
physical depletion. At still larger cylindrical distances, the profiles
are increasingly affected by the available Juno sampling geometry and
should not be interpreted as the intrinsic radial falloff of the main ring.

These radial features can be compared with boundaries previously identified
from imaging of Jupiter's ring system. Voyager observations placed the
transition between the main ring and halo near $\sim1.71~R_J$ and showed
the optically detectable halo extending inward toward Jupiter
\citep{showalter_jupiters_1987}. Galileo imaging subsequently located the
main-ring--halo transition near $122{,}500$~km ($1.716~R_J$) and traced
the equatorial halo inward to approximately $92{,}000\pm1{,}000$~km
($\sim1.29~R_J$) \citep{ockert-bell_structure_1999}. These characteristic
boundaries lie close to the $3{:}2$ and $2{:}1$ Lorentz resonances at
approximately $1.71$ and $1.40~R_J$, respectively, which have long been
associated with the radial organization of the Jovian halo
\citep{burns_lorentz_1985,horanyi_structure_1996,
horanyi_plasma_2010}. We note, however, that the present density profile
does not show a distinct feature at the $2{:}1$ resonance itself. Importantly, the apparent inner boundary of the halo in remote-sensing
observations is also limited by observational sensitivity.
\citet{pater_rings_2018} noted that inside the $2{:}1$ Lorentz resonance
the particle density becomes too low for ground-based detection, while
Galileo images reveal a faint extension of the halo closer to Jupiter.
Imaging this innermost population becomes increasingly difficult because of
strong telescopic stray light near Jupiter's limb. The Juno Waves
measurements provide a complementary in situ view of this region: the
cylindrical density profile remains clearly elevated inward of $1.4~R_J$
and extends to the innermost regions sampled by Juno. Thus, rather than
representing a sharp physical termination of the dust population, the
traditional optical inner boundary partly reflects the difficulty of
remotely detecting an increasingly tenuous population close to Jupiter.

\begin{figure*}[htb!]
    \centering
    \includegraphics[width=0.9\textwidth]
    {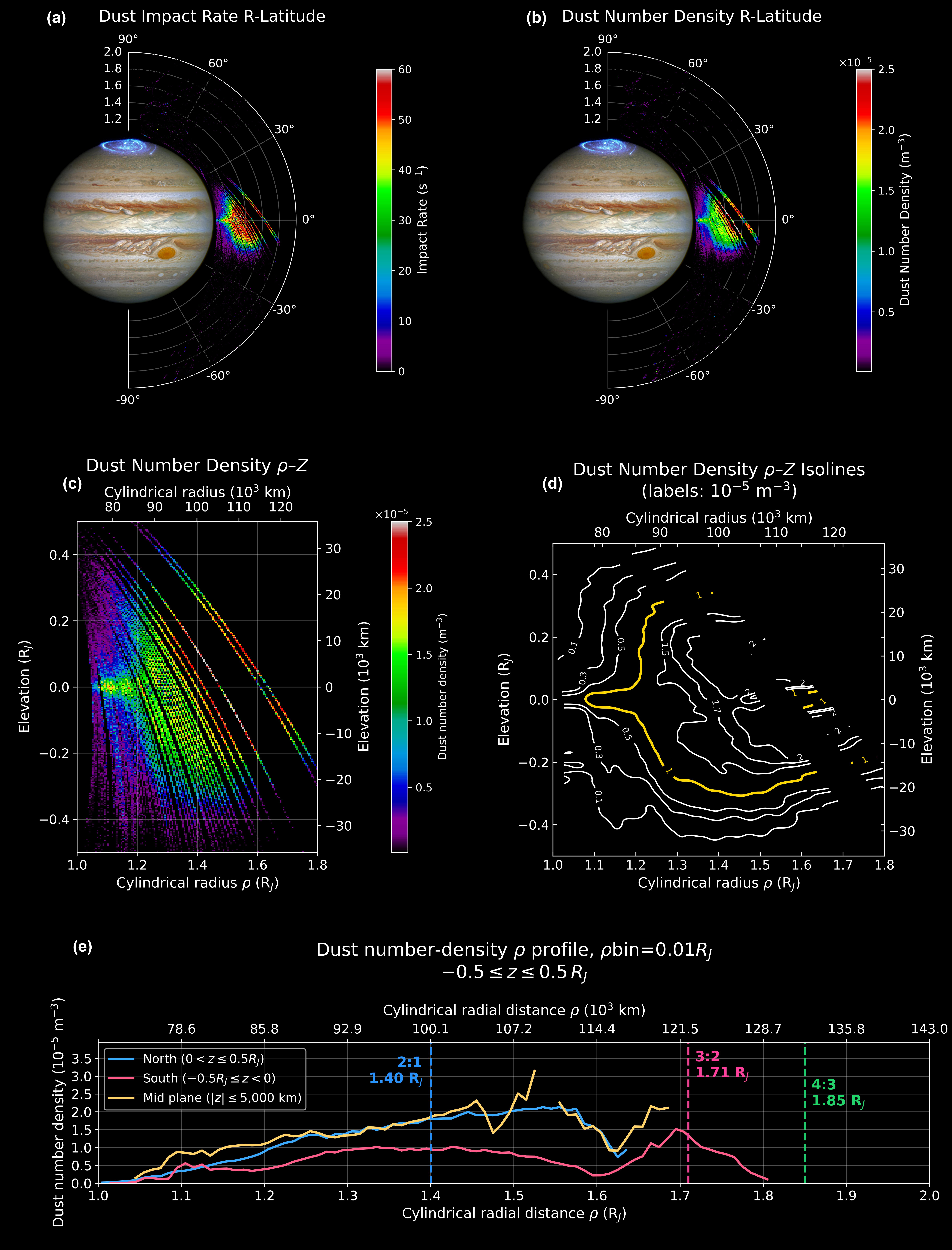}
    \caption{
    Spatial and radial structure of the dust population measured by
    Juno Waves.
    (a) Dust impact rate in $r$--latitude coordinates.
    (b) Inferred dust number density.
    (c) Dust number density in cylindrical $\rho$--$z$ coordinates.
    (d) Isodensity contours of the cylindrical distribution, with contour
    labels in units of $10^{-5}~{\rm m^{-3}}$.
    (e) Cylindrical radial number density profiles calculated with $\Delta\rho=0.01~R_J$ for the northern hemisphere ($0<z\leq0.5~R_J$; blue), the southern hemisphere ($-0.5~R_J\leq z<0$; pink), and the near-midplane region ($|z|\leq5{,}000$~km; yellow). Vertical dashed lines indicate the approximate locations of the $2{:}1$, $3{:}2$, and $4{:}3$ Lorentz resonances. The distributions show a compact inner equatorial population, strong vertical broadening through the halo region, and substantial radial and hemispheric structure.
    }
    \label{fig:dust_spatial_profiles}
\end{figure*}

\begin{figure*}[htb!]
    \centering
    \includegraphics[width=\textwidth]
    {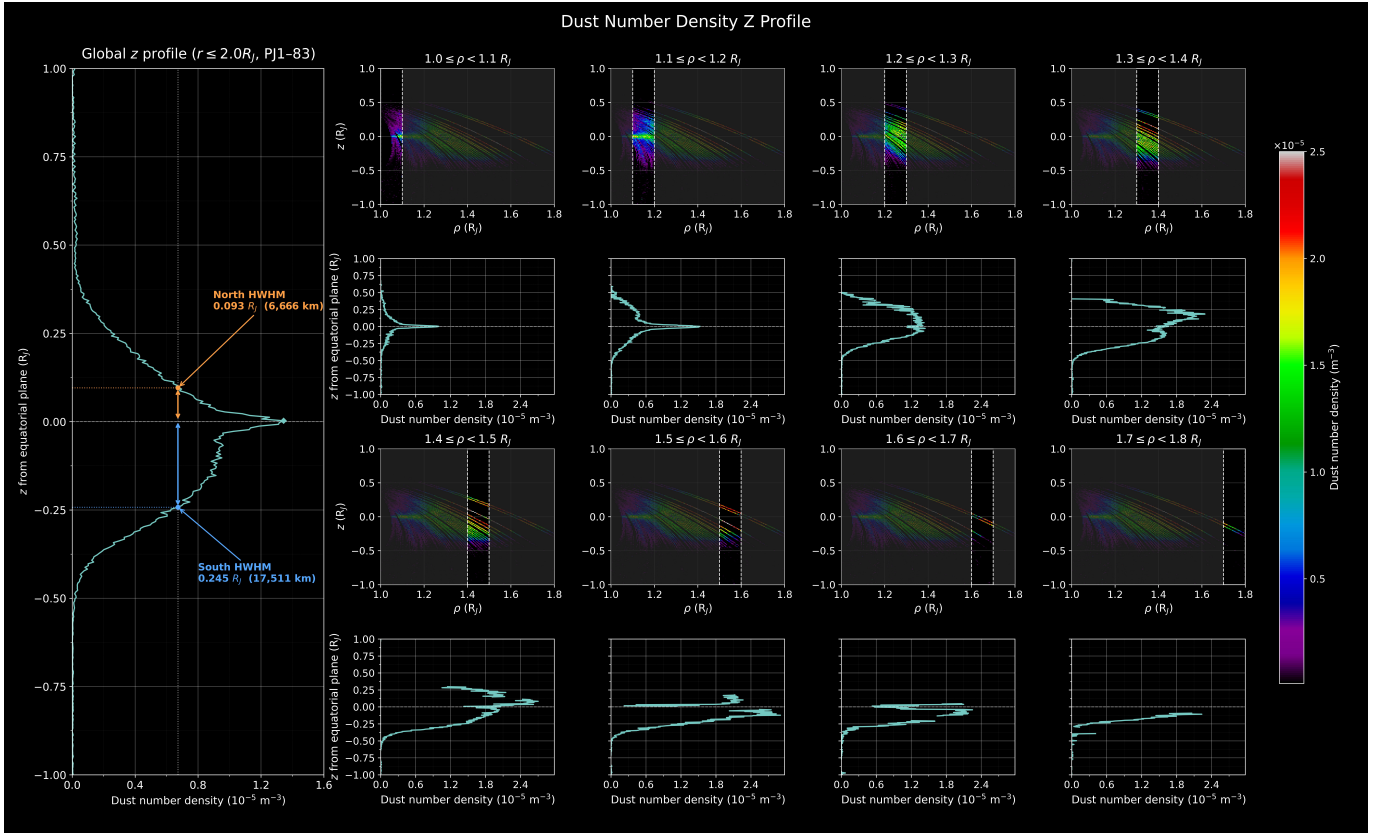}

    \caption{
    Radial evolution of the vertical dust number density distribution.
    The left panel shows the global $z$ profile for observations within
    $r\leq2.0~R_J$, with the northern and southern half-widths at
    half-maximum (HWHM) marked. The remaining panels show the
    $\rho$--$z$ number density distribution and corresponding vertical
    profiles in successive $0.1~R_J$ intervals from $\rho=1.0$ to
    $1.8~R_J$. The innermost population
    ($\rho\lesssim1.2~R_J$) is strongly concentrated near the equatorial
    plane and is approximately symmetric. Between approximately
    $1.2$ and $1.5~R_J$, the dust layer broadens and develops a stronger
    northward extension. At larger $\rho$, the apparent southern dominance
    is strongly influenced by the predominantly southern-hemisphere Juno
    sampling. The depletion near $z=0$ in several intermediate radial
    intervals coincides with changes in the Waves attenuation configuration.
    }
    \label{fig:dust_rhoz_profiles}
\end{figure*}

The hemispheric difference apparent in
Figure~\ref{fig:dust_spatial_profiles}e is examined further in
Figure~\ref{fig:dust_rhoz_profiles}, which shows the vertical dust
distribution as a function of cylindrical distance. The left panel presents
the global $z$ profile for observations within $r\leq2~R_J$, with the
northern and southern HWHM values marked. The remaining panels separate the
distribution into successive $\Delta\rho=0.1~R_J$ intervals, allowing the
radial evolution of the vertical structure to be examined directly.

The global profile is strongly peaked near $z=0$ but exhibits a pronounced
difference between the two hemispheres. The southern HWHM is
$0.245~R_J$ ($17{,}511$~km), compared with $0.093~R_J$
($6{,}666$~km) in the north. This global asymmetry, however, does not
represent a uniform southward enhancement at all radial distances. The
$\rho$-resolved profiles show that the hemispheric structure changes
systematically with cylindrical distance.

Within the innermost region,
$1.0\lesssim\rho\lesssim1.2~R_J$, the dust distribution is comparatively
symmetric about the equatorial plane. The highest densities remain strongly
concentrated near $z=0$, and the vertical width is small in both
hemispheres, consistent with the compact inner population identified in
Figure~\ref{fig:dust_spatial_profiles}.

A different morphology develops between approximately
$\rho=1.2$ and $1.5~R_J$. As the halo expands vertically, the northern side
first develops a larger HWHM than the southern side, and farther outward
within this interval the density at positive $z$ also becomes larger than
at corresponding negative $z$. The asymmetry therefore involves changes in
both the vertical width and density of the population rather than a simple
displacement of the density maximum. This northward development occurs over
the same radial interval in which the halo reaches its greatest vertical
extent near $\sim1.4~R_J$.

The detailed profiles close to the equatorial plane must be interpreted
together with the Waves attenuation effects discussed in
Section~\ref{sec:gain_effect}. Beginning near $\rho\sim1.2~R_J$, several
radial intervals exhibit an artificial reduction in inferred density around
$z=0$, corresponding to the attenuation-related star-shaped depletion in
the global dust maps. These local minima are therefore not interpreted as a
physical absence of equatorial dust. The broader envelopes and hemispheric
widths provide a more reliable indication of the vertical structure in this
radial range.

Beyond approximately $\rho=1.5~R_J$, the available measurements become
increasingly dominated by observations south of the equatorial plane.
The corresponding radial slices consequently show much larger densities at
negative $z$, but this apparent southern enhancement cannot be interpreted
as an intrinsic hemispheric asymmetry because comparable northern sampling
is largely absent. The global southward extension is therefore partly
influenced by the outer-region sampling geometry, while the north--south
structure is most robust inside approximately $1.5~R_J$, where both
hemispheres are substantially sampled.

The question of north--south symmetry in Jupiter's halo has a long
observational history. Early Voyager analysis reported a southward
displacement of the halo centroid of approximately 300~km
\citep{jewitt_jovian_1981}. Reanalysis by
\citet{showalter_jupiters_1987}, however, found much smaller vertical
offsets, comparable to the image resolution, and showed that much of the
earlier apparent displacement resulted from image-background processing.
The large-scale halo was therefore concluded to be approximately symmetric
about Jupiter's equatorial plane. Galileo imaging subsequently reached a
similar conclusion, with \citet{ockert-bell_structure_1999} reporting no
obvious north--south asymmetry in the global halo morphology.

The optical and in situ measurements probe different aspects of the dust
distribution. Remote-sensing observations integrate scattered light along
the line of sight and naturally average over local spatial and temporal
structure, whereas Juno Waves samples the dust population locally along the
spacecraft trajectory. Using the first 21 Juno ring-plane crossings,
\citet{ye_juno_2020} found that the fitted center of the inner dust layer
varied by several hundred kilometers from one perijove to another, including
offsets of approximately $+244$ and $+236$~km during PJ3 and PJ4,
$-375$~km during PJ5, and nearly $+496$~km during PJ13. These measurements
demonstrate that the locally sampled dust layer need not remain exactly
centered on the System~III equatorial plane even if the large-scale,
time-averaged halo is approximately symmetric.

The extended Juno Waves data set reveals a related but distinct form of
vertical organization. Rather than showing a single persistent displacement
of the dust layer toward one hemisphere, the observations reveal a
systematic radial evolution of its hemispheric structure. The innermost
population is approximately symmetric, whereas a northward enhancement
develops as the halo expands between approximately $1.2$ and $1.5~R_J$.
At still larger cylindrical distances, the present sampling becomes too
strongly biased toward the south for a comparable hemispheric assessment.
The observed asymmetry is therefore radius dependent rather than a simple
global north--south offset.

A possible geometrical reference for this transition is provided by
Jupiter's magnetic equator. Using the JRM33 internal field \citep{connerney_new_2022} together with
the Con2020 external magnetodisc field \citep{connerney_jovian_2020}, we calculated the longitude-averaged magnetic
equator using both $B_r=0$ and $B_\rho=0$ definitions
(Appendix~\ref{app:dust_dynamics_support}). The two definitions produce
very similar radial trends. At small cylindrical distances, the mean
magnetic equator lies south of the System~III equatorial plane, crosses the equatorial plane near $\rho\sim1.23$--$1.24~R_J$, and then shifts progressively
northward with increasing cylindrical distance. The onset of this northward
magnetic-equator displacement occurs close to the radial range in which the
observed dust layer begins to develop its strongest northward broadening.
The correspondence is therefore robust to the particular magnetic-equator
definition adopted. The magnetic-equator displacement itself is much smaller than the observed
halo thickness. Near $1.4~R_J$, for example, its magnitude is only of order
$0.02~R_J$, compared with a halo HWHM of approximately $0.367~R_J$.
The magnetic equator therefore cannot by itself account for either the
large vertical extent or the hemispheric structure of the dust population.
Instead, its radial displacement provides a possible symmetry-breaking
reference geometry for the Lorentz forcing of charged grains.

The response of the dust population to this magnetic geometry also depends
strongly on grain size and on the relative ordering of charging, dynamical,
and Jovian-rotation timescales. The test-particle calculations described in
Appendix~\ref{app:dust_dynamics_support} show that for
$a\lesssim0.3~\mu$m the simulated mean lifetime is shorter than one
System~III rotation period ($\sim9$~h~55~min) and is comparable to or
shorter than the time required to approach local charging equilibrium.
These small grains therefore occupy a rapid-loss,
nonequilibrium-charging regime. Although their large charge-to-mass ratios
allow strong instantaneous electromagnetic forcing, they can be removed
before sampling a complete rotation of Jupiter's non-axisymmetric magnetic
field, making sustained phase-coherent Lorentz-resonant evolution unlikely.

For grains with $a\gtrsim0.4~\mu$m, the charging time becomes much shorter
than the dynamical lifetime, while the lifetime extends over many Jovian
rotations. These particles can therefore interact repeatedly with the
rotating magnetic-field structure after approaching local charging
equilibrium, making long-term electromagnetic transport and
Lorentz-resonant effects more dynamically relevant. The transition between
these regimes indicates that the same electromagnetic transport mechanism
should not necessarily be expected to operate uniformly across the full
grain-size distribution.

Taken together, the Juno Waves observations reveal a dust population with
substantial radial, vertical, and hemispheric organization. A compact,
approximately symmetric equatorial population persists inward of the
optically observed halo. Farther outward, the dust layer expands rapidly,
develops a northward hemispheric structure between approximately
$1.2$ and $1.5~R_J$, and subsequently contracts toward the main-ring
region. The radial density distribution contains broad halo-related
structure and remains measurable well inward of the traditional optical
inner boundary, demonstrating that this boundary does not represent a sharp
termination of the dust population. The radial evolution of Jupiter's
magnetic geometry and the strongly size-dependent charging and dynamical
timescales provide physical context for the observed vertical organization,
although neither alone provides a quantitative explanation. A full
interpretation will require three-dimensional charged-dust dynamics with
time-dependent charging and the actual Juno sampling geometry.

\subsection{Global Distribution of Solitary Waves}
\label{sec:solitary_distribution}

\begin{figure*}[htb!]
    \centering
    \includegraphics[width=\textwidth]{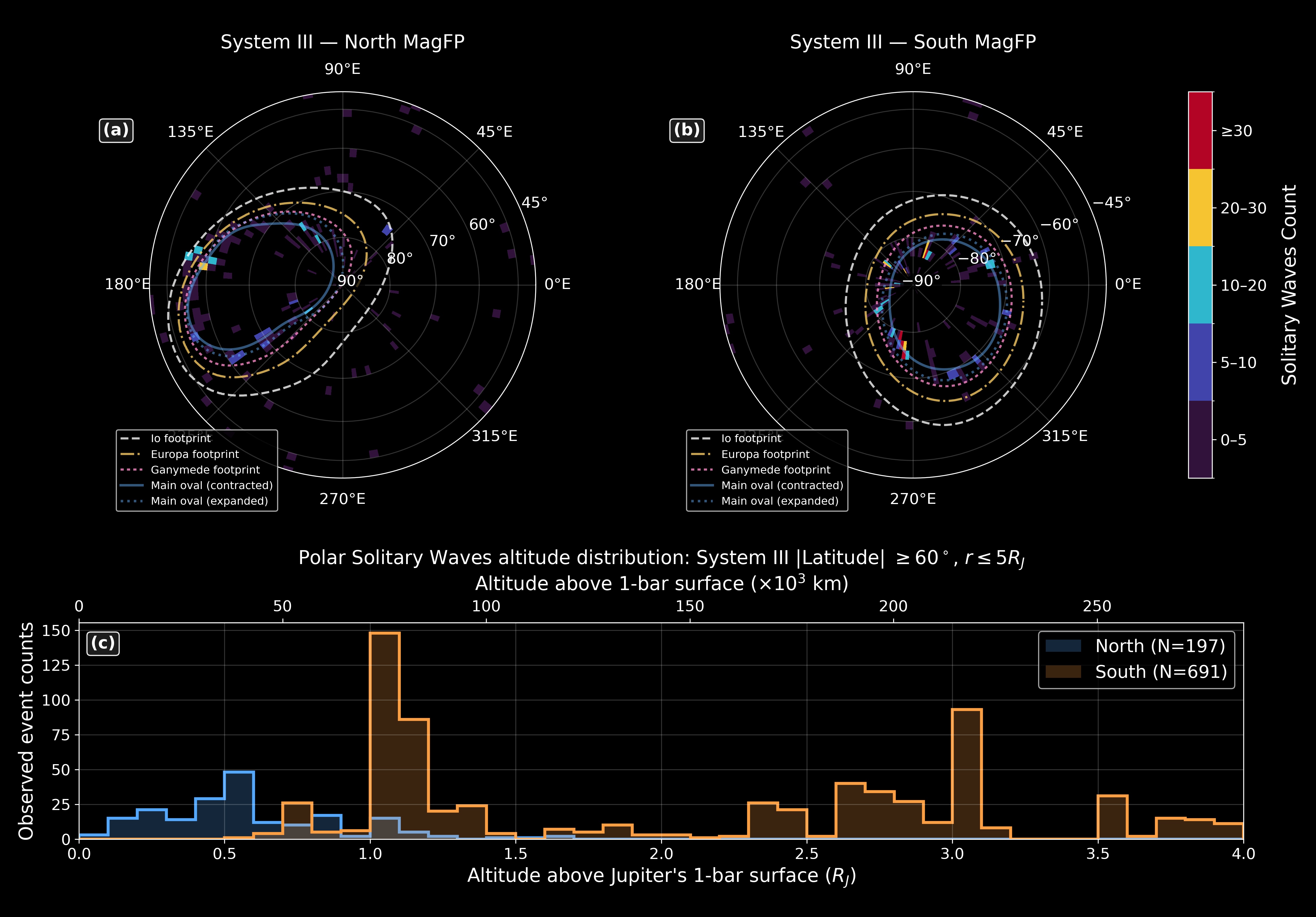}
    \caption{
    Magnetic-footprint and altitude distributions of high-latitude
    Solitary Waves detected by Juno Waves.
    (a--b) System~III magnetic footprints of the field lines associated
    with Solitary Wave observations in the northern and southern
    hemispheres, respectively. Colors indicate the number of detected
    events in each spatial bin. The magnetic footprints of Io, Europa,
    and Ganymede are shown for reference, together with the observed
    location of Jupiter's main auroral oval from \citet{bonfond_tails_2017}.
    The highest-count Solitary Wave regions show a strong spatial
    correspondence with these auroral structures, particularly the
    main oval.
    (c) Altitude distribution of high-latitude
    ($|\mathrm{latitude}|\geq60^\circ$) Solitary Wave observations within
    $r\leq5~R_J$ for the northern ($N=197$) and southern ($N=691$)
    hemispheres. Altitude is measured above Jupiter's 1-bar surface.
    }
    \label{fig:solitary_distribution}
\end{figure*}

Electrostatic solitary waves (ESWs) are short-duration, nonlinear electric
field structures that have been observed in a variety of space plasma
environments. At Earth, broadband electrostatic noise (BEN) was later shown
to consist, at least in part, of sequences of bipolar solitary structures
associated with electron phase-space holes
\citep{matsumoto_electrostatic_1994}. Similar electrostatic solitary structures have
subsequently been reported in other plasma environments, including the
lunar wake and regions above lunar magnetic anomalies
\citep{hashimoto_electrostatic_2010}.

At Jupiter, the observational history of these structures extends back to
Voyager. Using Voyager~1 plasma-wave observations,
\citet{barbosa_broadband_1981} identified impulsive BEN in Jupiter's middle
magnetosphere, primarily near the edges of the plasma sheet between
approximately $10$ and $30~R_J$. These emissions were associated with
large-scale field-aligned current systems connected to Jupiter's auroral
region. Subsequent examination of Voyager wideband waveforms showed that
at least some of the Jovian BEN contains electrostatic solitary structures
similar to those observed in terrestrial auroral regions
\citep{kurth_juno_2017}. Juno Waves has since enabled these short-duration
structures to be observed directly in Jupiter's polar magnetosphere.
More recent Juno observations have shown that large-amplitude solitary
waves are an important component of Jupiter's auroral plasma environment,
including regions associated with field-aligned currents and particle
acceleration \citep{elliott_generation_2020,sulaiman_jupiters_2022}.

The extended Juno Waves data set analyzed here allows us to move beyond
individual events and examine the large-scale spatial organization of the
Solitary Wave population. In contrast to the strong equatorial
concentration of Dust Impacts, Solitary Waves are observed predominantly
at high magnetic latitudes. To investigate their relationship with
Jupiter's auroral and magnetospheric structure, we trace the magnetic field
line associated with each high-latitude Solitary Wave observation to
Jupiter's northern or southern ionosphere. The resulting System~III
magnetic footprints are shown in
Figure~\ref{fig:solitary_distribution}a--b. For comparison, the magnetic
footprints of Io, Europa, and Ganymede are overplotted together with the
observed location of Jupiter's main auroral oval
\citep{bonfond_tails_2017}.

The most striking feature of the footprint distributions is their strong
spatial correspondence with these auroral reference structures. In both
hemispheres, nearly all bins containing more than 10 Solitary Wave events
are located on or close to the main auroral oval or the satellite
footprints. The correspondence is particularly strong with the main oval,
where many of the highest-count bins are concentrated. This result
indicates that Solitary Waves are preferentially observed on magnetic field
lines connected to Jupiter's auroral structures rather than being randomly
distributed throughout the polar region.

This spatial association is consistent with previous observations and
proposed generation mechanisms for electrostatic solitary structures.
Solitary waves have previously been reported in Jupiter's auroral regions,
including regions associated with field-aligned currents, electron beams,
and particle acceleration
\citep{kurth_juno_2017,elliott_generation_2020,sulaiman_jupiters_2022}. The strong correspondence
found here between the statistically identified Solitary Wave population
and the auroral reference structures therefore provides additional
observational evidence that their occurrence is closely related to the
plasma and current systems associated with Jupiter's auroral magnetosphere.

The magnetic footprints in Figure~\ref{fig:solitary_distribution}a--b
represent the ionospheric mapping of the magnetic field lines on which the
Solitary Waves were observed, rather than the physical locations at which
the waves were necessarily generated. The correspondence with the main
oval therefore indicates that the waves are preferentially observed on
field lines connected to Jupiter's main auroral region, but does not by
itself determine their generation altitude or source location.

Figure~\ref{fig:solitary_distribution}c shows the altitude distribution of
the high-latitude Solitary Wave observations
($|\mathrm{latitude}|\geq60^\circ$, $r\leq5~R_J$). A total of 197 events
are identified in the northern hemisphere and 691 in the southern
hemisphere. The northern detections are concentrated primarily at lower
altitudes, with most events observed below approximately $1~R_J$ above
Jupiter's 1-bar surface. In contrast, the southern detections span a broader
altitude range, with a strong concentration near $\sim1~R_J$ and additional
detections extending to approximately $4~R_J$. This north--south difference
is partly influenced by Juno's orbital geometry, which samples substantially
higher altitudes over the southern auroral region than over the northern
auroral region.

Together, the magnetic-footprint and altitude distributions demonstrate
that the Solitary Waves constitute a high-latitude plasma population
spatially distinct from the equatorially concentrated Dust Impacts. Most
importantly, the concentration of high-count events along or near the main
auroral oval and satellite footprints reveals a strong spatial association
between the Solitary Wave population and Jupiter's auroral magnetic
structure. This statistical relationship is consistent with previous
observations of solitary structures in Jovian auroral current regions and
supports a close connection between these nonlinear plasma structures and
the field-aligned current and particle acceleration processes operating in
Jupiter's auroral magnetosphere \citep{barbosa_broadband_1981, kurth_juno_2017,elliott_generation_2020,sulaiman_jupiters_2022}.

\section{Discussion}
\label{sec:Discussion}

\subsection{Effects of Waves Gain and Attenuation Settings}
\label{sec:gain_effect}

Before interpreting the spatial and size-dependent structure of the dust
population, it is necessary to consider the influence of the Juno Waves
receiver configuration on dust detection. For the LFR\_LO\_E channel used in
this study (50~Hz--20~kHz), the effective dynamic range is controlled by
multiple cascaded stages, including selectable attenuation in the electric
preamplifier, attenuation within the Low Frequency Receiver (LFR), and a
programmable gain amplifier in the LFR electric-field channel
\citep{kurth_juno_2017}. The electric preamplifier provides discrete
attenuation levels including 19.0 and 25.3~dB. Different combinations of
these settings allow Waves to accommodate the large range of electric-field
amplitudes encountered along the Juno trajectory while reducing the
occurrence of receiver clipping.

The attenuation configuration determines the range of dust-impact pulse
amplitudes that can be reliably observed. Lower attenuation provides greater
sensitivity to weak pulses, but sufficiently large signals can approach the
upper dynamic-range limit and become clipped. Stronger attenuation extends
the measurable range toward larger-amplitude signals, but weak impact pulses
can fall below the effective detection threshold. Thus, different receiver
configurations preferentially sample different portions of the underlying
dust population. Similar gain-dependent selection effects are well
documented in Cassini RPWS dust measurements
\citep{kurth_cassini_2006,ye_properties_2014}. For Juno Waves,
\citet{ye_juno_2020} showed that using only events detected in the
lowest-gain state (attenuation on) produced a voltage pulse-height
distribution with a slope of approximately $-1.3$, corresponding to an
inferred grain-size distribution slope of approximately $-3.9$. In
contrast, using all unclipped impacts while accounting for the receiver
attenuation yielded a steeper differential voltage-distribution slope of
approximately $-1.7$, corresponding to a grain-size distribution slope of
approximately $-5.1$. They noted that the lowest-gain selection favors
larger impacts and therefore preferred the distribution obtained using
both attenuation settings. An additional uncertainty arises from the location of an impact relative to
the Waves antennas. The amplitude and shape of a dust-impact signal can
depend on the impacted spacecraft element and its coupling to the antennas
\citep{ye_properties_2014,ye_understanding_2019,ye_juno_2020}. Because Juno
Waves detects impacts on the spacecraft body and solar panels as well as on
the antenna elements, variations in impact location may introduce additional
uncertainty into the conversion from pulse amplitude to grain size and,
consequently, the inferred grain-size index.

The peak electric-field amplitude of a dust-impact signal can be used to
estimate the grain size. In the reference calculation used throughout this
study, we relate the measured peak electric field to the collected
impact-generated charge through

\begin{equation}
\alpha Q_{\rm imp}
=
C_{\rm eff}L_{\rm eff}E_{\rm peak},
\label{eq:impact_charge_epeak}
\end{equation}

where $C_{\rm eff}=46.69$~pF is the effective capacitance, $L_{\rm eff}=0.5$~m is the effective antenna length, and
$\alpha$ is the charge-coupling efficiency. We adopt $\alpha=1$ for the
reference calculation. The total impact-generated charge is related to the
dust mass using a simplified form of the empirical impact-ionization
relation,

\begin{equation}
Q_{\rm imp}
=
\gamma m_{\rm d}v_{\rm rel}^{3.5},
\label{eq:impact_charge_mass}
\end{equation}

where $\gamma=0.7$. This approximation follows the commonly used
aluminum-target scaling of \citet{mcbride_meteoroid_1999}, as summarized and
discussed by \citet{collette_micrometeoroid_2014}. Assuming spherical
silicate grains with density
$\rho_{\rm d}=3000~{\rm kg\,m^{-3}}$, the inferred grain radius is

\begin{equation}
a =
\left[
\frac{3C_{\rm eff}L_{\rm eff}E_{\rm peak}}
{4\pi\rho_{\rm d}\alpha\gamma v_{\rm rel}^{3.5}}
\right]^{1/3}.
\label{eq:grain_radius}
\end{equation}

\begin{figure*}[htb!]
    \centering
    \includegraphics[width=0.7\textwidth]
    {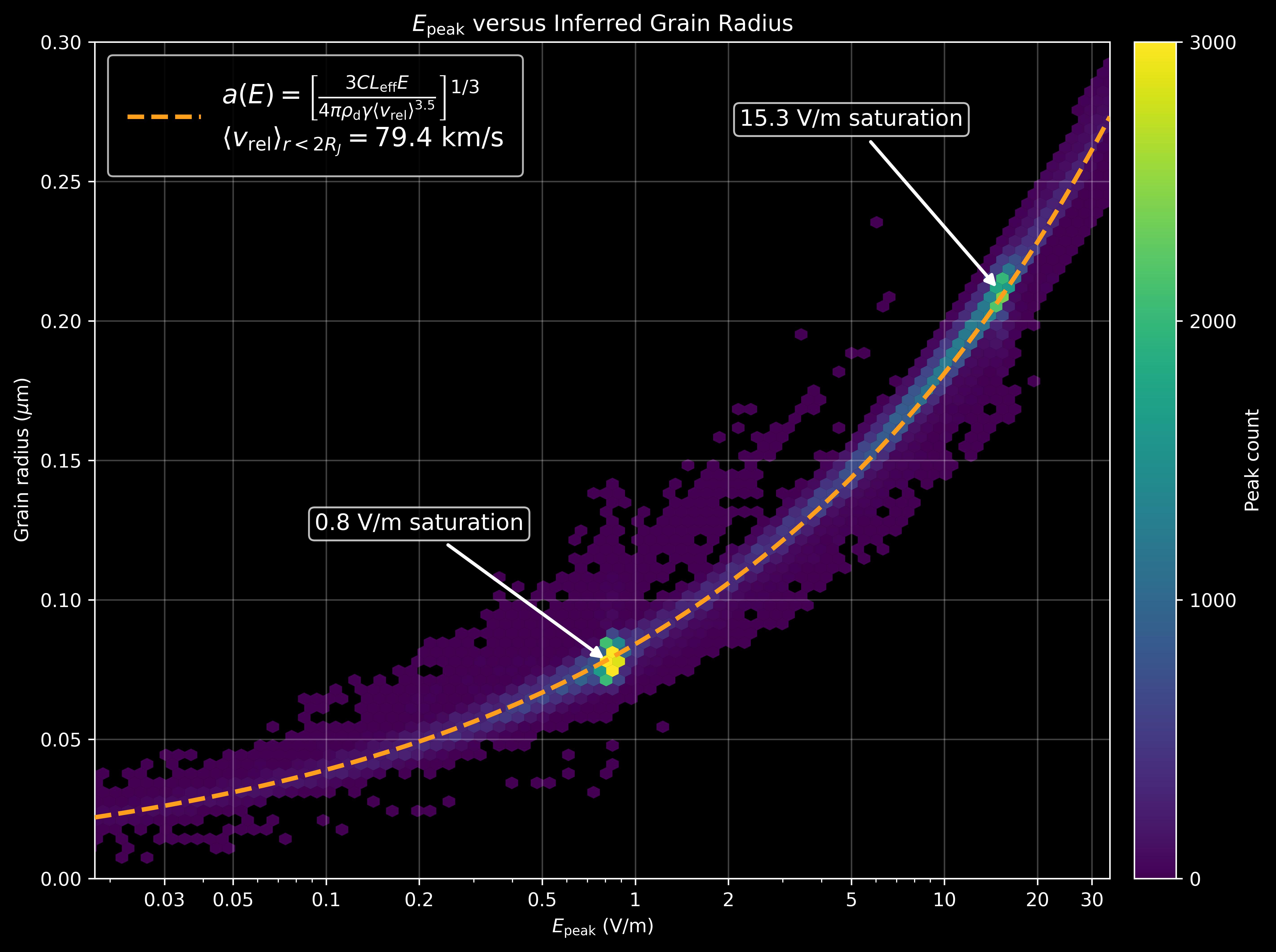}

    \caption{
    Peak electric-field amplitude, $E_{\rm peak}$, versus inferred dust grain
    radius. Color indicates the number of events in each bin. The dashed
    curve shows the relation given by Equation~(\ref{eq:grain_radius})
    evaluated using the reference parameters and the global mean relative
    impact velocity
    $\langle v_{\rm rel}\rangle=79.4~{\rm km\,s^{-1}}$.
    The strong concentrations near 0.8 and $15.3~{\rm V\,m^{-1}}$
    correspond to upper-amplitude limits associated with different Waves
    gain/attenuation configurations and should not be interpreted as
    distinct physical grain-size populations.
    }
    \label{fig:dust_epeak_grainsize}
\end{figure*}

Figure~\ref{fig:dust_epeak_grainsize} shows the measured peak electric-field
amplitude as a function of inferred grain radius. Because both $E_{\rm peak}$
and $v_{\rm rel}$ enter the grain-size estimate, the dashed curve is shown
for reference using the global mean relative impact velocity,
$\langle v_{\rm rel}\rangle=79.4~{\rm km\,s^{-1}}$. Two pronounced
concentrations occur near $E_{\rm peak}=0.8$ and
$15.3~{\rm V\,m^{-1}}$. Rather than representing two preferred physical
grain populations, these concentrations correspond to upper-amplitude limits
associated with different Waves attenuation configurations. As the
recorded pulse approaches the measurable limit of a given configuration,
larger physical amplitudes cannot be recovered accurately, causing events to
accumulate near these values. The instrumental limits therefore propagate
directly into the inferred grain-radius distribution.

In addition to these instrumental limits, the absolute grain-size scale
depends on the parameters adopted in Equation~(\ref{eq:grain_radius}).
In particular,

\begin{equation}
a \propto
\left(
\frac{C_{\rm eff}}{\alpha\gamma}
\right)^{1/3},
\label{eq:grain_radius_scaling}
\end{equation}

so uncertainties in $C_{\rm eff}$, $\alpha$, and $\gamma$ directly affect
the inferred radius. Our reference calculation adopts
$C_{\rm eff}=46.69$~pF, $\alpha=1$, and $\gamma=0.7$, and produces a
grain-radius distribution dominated by particles of order
$\sim0.2~\mu{\rm m}$. These values should be regarded as reference
parameters rather than a unique absolute calibration of the dust size.

The coupling efficiency $\alpha$ is particularly difficult to determine
because it depends on how the impact-generated plasma couples to the
spacecraft and antenna system. In their analysis of Voyager plasma-wave
dust detections, \citet{gurnett_micron-sized_1983} estimated that $\alpha$ could plausibly
range from approximately 0.1 to 1, with a nominal value near 0.6. Although
these values cannot be directly transferred to Juno, they illustrate the
magnitude of the coupling uncertainty. For example, reducing $\alpha$ from
1 to 0.1 increases the inferred radius by a factor of
$10^{1/3}\simeq2.15$, such that a grain inferred to have a radius of
$0.20~\mu{\rm m}$ in the reference calculation would instead correspond to
approximately $0.43~\mu{\rm m}$.

The effective capacitance and impact-ionization coefficient introduce
additional systematic uncertainties. The value
$C_{\rm eff}=46.69$~pF is adopted here as the reference value for the
voltage-to-charge conversion, but the effective value appropriate for a
dust impact depends on the coupling between the impact plasma, spacecraft,
and Waves antenna system. Similarly, $\gamma$ depends on the impact
ionization efficiency and therefore on quantities such as projectile and
target composition and impact conditions
\citep{collette_micrometeoroid_2014}. Because the inferred radius varies as
$C_{\rm eff}^{1/3}$ and $\gamma^{-1/3}$, uncertainties in these parameters
can further shift the absolute grain-size scale. 

This systematic uncertainty is also illustrated by the previous Juno dust
analysis of \citet{ye_juno_2020}. Using an empirical relation of the form
$V=Kmv^{3.5}$ and scaling from previous plasma-wave dust measurements, they
estimated a characteristic grain radius of approximately
$1.4~\mu{\rm m}$ for the Juno-detected dust population. The difference
between this estimate and the $\sim0.2~\mu{\rm m}$ characteristic scale
obtained using our reference parameters emphasizes that the absolute grain
size remains strongly model dependent. We therefore use the inferred radii
primarily as a consistent reference scale for examining relative spatial
variations rather than as a precise absolute measurement of particle size.

These upper-amplitude limits also prevent a precise determination of the
sizes of grains whose signals approach saturation. At the opposite end of
the amplitude range, stronger attenuation preferentially removes weak pulses,
introducing an additional selection against smaller impacts. Taken together,
the receiver response and the uncertainties in $C_{\rm eff}$, $\alpha$, and
$\gamma$ affect the absolute grain-size distribution. The relative spatial
organization obtained by applying the same reference conversion throughout
the data set is less sensitive to these absolute calibration uncertainties.

\begin{figure}[htb!]
    \centering
    \includegraphics[width=0.6\textwidth]
    {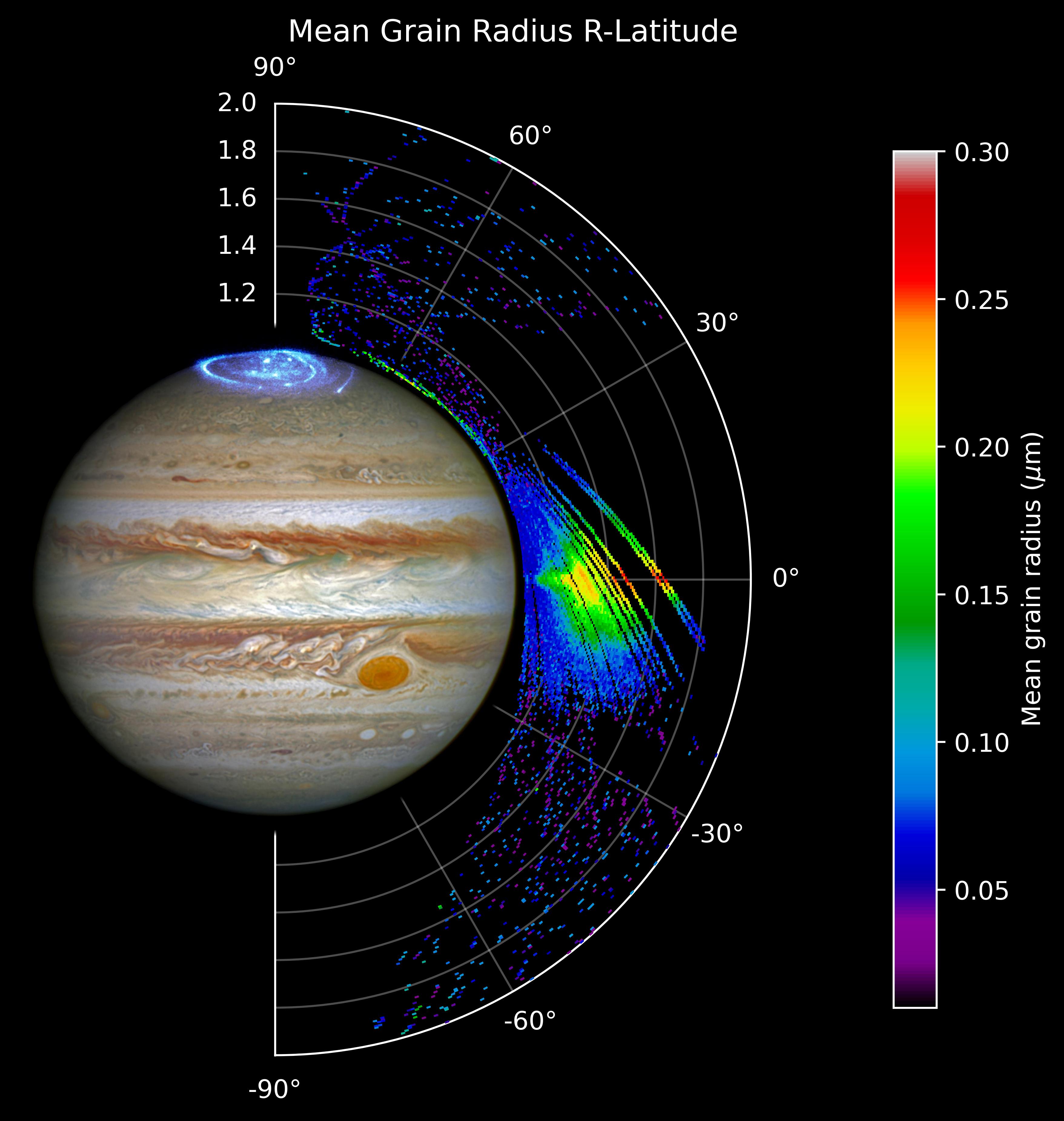}

    \caption{
    Spatial distribution of the mean inferred dust grain radius in radial
    distance--latitude coordinates. High-latitude detections are dominated
    by relatively small inferred grains. Within the low- and middle-latitude
    population, the inferred grain radius generally increases with radial
    distance toward the Jovian halo region. The values shown are calculated
    using the reference parameters adopted in
    Equation~(\ref{eq:grain_radius}); absolute grain sizes remain subject to
    uncertainties in the receiver response, impact ionization, and
    spacecraft--antenna coupling.
    }
    \label{fig:dust_grainsize_spatial}
\end{figure}

Despite these limitations, the spatial distribution of inferred grain radius
shows systematic organization
(Figure~\ref{fig:dust_grainsize_spatial}). High-latitude dust detections are
dominated by relatively small inferred grains, typically with reference
radii below $\sim0.1~\mu{\rm m}$, and the number of detected impacts in
these regions is also substantially smaller than at low latitudes. Within
the dominant low- and middle-latitude population, the inferred grain radius
generally increases with radial distance, with larger grains becoming more
common toward the Jovian halo region.

This size-dependent spatial organization is qualitatively consistent with
models of charged-dust transport in Jupiter's inner magnetosphere. Smaller
grains have larger charge-to-mass ratios and are therefore more strongly
affected by electromagnetic forces, allowing them to acquire larger orbital
inclinations, whereas larger grains remain more closely confined to the
equatorial region \citep{horanyi_plasma_2010}. The predominance of smaller
reference grain sizes at high latitudes is consistent with this picture.
The increase in inferred grain size with radial distance within the low- and
middle-latitude population is also broadly consistent with size-dependent
transport within the Jovian ring and halo system. Given the uncertainty in
the absolute grain-size scale, however, these comparisons should be regarded
primarily in terms of relative spatial trends.

\begin{figure}[htb!]
    \centering
    \includegraphics[width=0.7\textwidth]
    {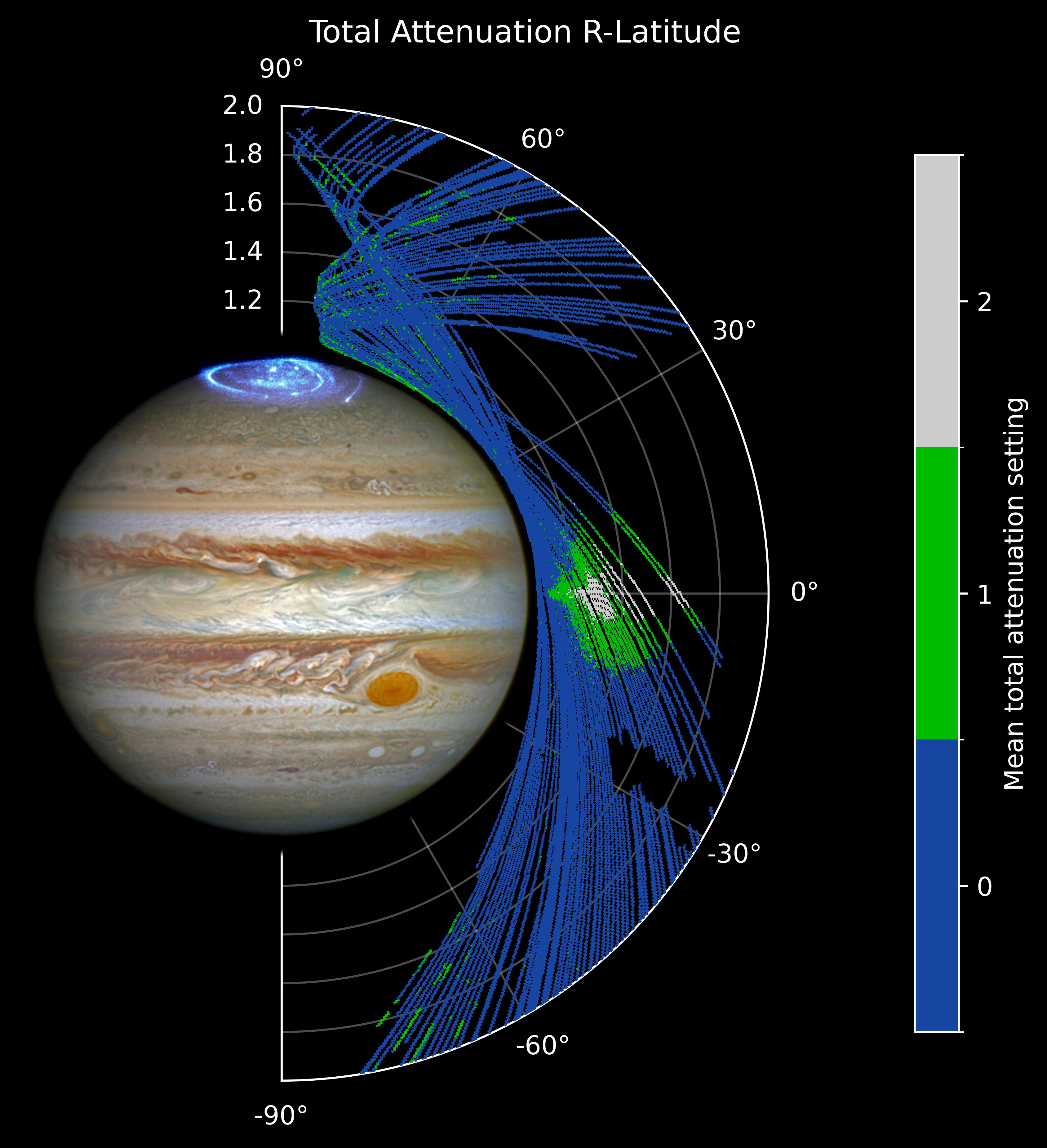}

    \caption{
    Mean total Waves attenuation along the Juno trajectory in
    radial-distance--latitude coordinates. The attenuation structure near
    the equatorial region closely corresponds to the star-shaped depletion
    in the detected dust population between approximately
    $1.2$ and $1.4~R_J$. Increased attenuation reduces sensitivity to weak
    impact pulses, indicating that this localized feature is primarily
    instrumental rather than a physical dust cavity.
    }
    \label{fig:waves_attenuation}
\end{figure}

The influence of the receiver configuration is also directly visible in the
spatial distribution of detected impacts. Figure~\ref{fig:waves_attenuation}
shows the mean total Waves attenuation encountered along the Juno trajectory
in radial-distance--latitude coordinates. The attenuation structure near the
equatorial plane closely follows the conspicuous star-shaped depletion in the
dust distributions between approximately $1.2$ and $1.4~R_J$ shown in
Figure~\ref{fig:dust_spatial_profiles}. Regions with stronger attenuation
have reduced sensitivity to weak impact pulses, naturally producing fewer
detected events. The close spatial correspondence therefore indicates that
the star-shaped feature is predominantly an instrumental signature rather
than a physical cavity in the Jovian dust population.

A second example occurs in the cylindrical radial profile discussed in
Section~\ref{sec:dust_global_distribution}. The pronounced minimum near
$\rho\sim1.6~R_J$ coincides with a substantial change in the Waves
attenuation configuration. We therefore do not interpret this minimum as an
intrinsic depletion of the dust population. In contrast, broader structures
that persist across regions observed under different receiver configurations,
such as the extended halo distribution and the increase in dust density
toward the main-ring--halo transition near $\rho\sim1.7~R_J$, are less
readily explained by this instrumental effect.

Overall, the Waves gain and attenuation configuration affects the dust
analysis at both ends of the measurable pulse-amplitude range. Saturation and
clipping limit the recovery of the largest impact amplitudes, whereas
stronger attenuation reduces sensitivity to weak pulses and preferentially
removes smaller impacts. In addition, uncertainties in $C_{\rm eff}$,
$\alpha$, and $\gamma$ affect the absolute conversion from pulse amplitude
to grain radius. The inferred grain sizes should therefore be regarded as a
reference scale rather than a precise absolute calibration. In contrast,
the relative radial and latitudinal variations are more robust because the
same conversion is applied consistently throughout the data set.
Identifying these instrumental and calibration effects is essential for
separating receiver-induced features from the larger-scale physical
structure of Jupiter's inner dust environment.

\section{Summary}
\label{sec:summary}

We have used nearly a decade of Juno Waves LFR\_LO\_E burst observations,
spanning PJ01--PJ83, to investigate the spatial distributions of dust impacts
and transient plasma structures in Jupiter's innermost magnetosphere. A
two-stage CNN framework was developed to first identify transient
electric-field waveforms and then separate them into Dust Impact, Solitary
Wave, and Other Waves. This approach enables the dust population to be
mapped over the full Juno data set while preserving physically distinct
nondust transients for independent investigation.

The dust-impact population is strongly concentrated near Jupiter's
equatorial plane but shows substantial radial and vertical structure. Inside
$\sim1.2~R_J$, the dust forms a compact equatorial layer with
characteristic vertical HWHM of only $\sim0.03$--$0.08~R_J$, and localized
density enhancements are present near $\rho\sim1.10$ and $1.15~R_J$, although their inferred number densities are more uncertain because these regions coincide with relatively small effective spacecraft impact areas. Moving outward, the layer broadens rapidly and reaches its maximum observed
vertical extent near $\sim1.4~R_J$, where the HWHM is approximately
$0.367~R_J$ ($\sim2.6\times10^4$~km). It then narrows again toward the
main-ring region, producing the expansion--contraction morphology expected
for a toroidal halo. The measured vertical extent is comparable to the
diffuse outer boundaries inferred from Voyager and Galileo imaging.

The cylindrical radial density profile further reveals a structured,
nonmonotonic dust distribution. A broad elevated-density region extends over
approximately $\rho\sim1.3$--$1.5~R_J$, with the $2{:}1$ Lorentz resonance
lying within this plateau rather than corresponding to a distinct density
feature. An enhancement in the cylindrical profile occurs at
$\rho=1.695~R_J$, where the vertically averaged density reaches
$1.52\times10^{-5}~{\rm m^{-3}}$. This location lies close to both the
main-ring--halo transition and the $3{:}2$ Lorentz resonance at
$\sim1.71~R_J$. Importantly, Juno Waves continues to detect an elevated dust
population inward of the traditional optical halo boundary near
$\sim1.4~R_J$, demonstrating that the apparent inner edge seen remotely
does not correspond to a sharp termination of the dust population.

The vertical distribution also exhibits hemispheric structure. The global
profile has a southern HWHM of $0.245~R_J$, compared with
$0.093~R_J$ in the north. However, the $\rho$-resolved profiles demonstrate
that this asymmetry is strongly radius dependent: the innermost
$\rho\lesssim1.2~R_J$ population is approximately symmetric, whereas a
northward broadening and density enhancement develops between
$\sim1.2$ and $1.5~R_J$. At larger cylindrical distances the available
observations are dominated by southern-hemisphere sampling. The onset of the
northward enhancement occurs near the radial distance at which the
longitude-averaged magnetic equator calculated from JRM33 and Con2020 moves
north of the System~III equatorial plane, suggesting that the magnetic-field
geometry may contribute to the observed hemispheric organization.

The interpretation of smaller-scale structures is limited by changes in the
Waves gain and attenuation configuration. Instrumental upper-amplitude limits
produce strong concentrations near $E_{\rm peak}=0.8$ and
$15.3~{\rm V\,m^{-1}}$ and limit quantitative grain-size estimates, while
stronger attenuation reduces sensitivity to weak impact pulses. The
star-shaped depletion observed near $1.2$--$1.4~R_J$ and the minimum near
$\rho\sim1.6~R_J$ coincide with changes in receiver attenuation and are
therefore interpreted primarily as instrumental rather than physical dust
depletions. After accounting for these effects, the larger-scale halo
structure, radial enhancements, and vertical evolution remain robust
features of the Juno observations.

Finally, the multi-class analysis reveals a Solitary Wave population that is
spatially distinct from the Dust Impacts and occurs predominantly at high
latitudes. The magnetic footprints of these events show a strong spatial
correspondence with Jupiter's auroral structures: nearly all high-count
regions ($>10$ events per bin) lie on or close to the main auroral oval or
the Io, Europa, and Ganymede footprints, with the strongest overall
correspondence occurring along the main oval. This organization is
consistent with previous observations of solitary structures in Jupiter's
auroral current regions and supports a close connection between the
Solitary Wave population and the field-aligned current and particle
acceleration processes operating in Jupiter's auroral magnetosphere.
Solitary Waves are detected over a broad range of polar altitudes, although
the observed north--south difference in altitude coverage is partly
influenced by Juno's orbital sampling geometry.

Together, these observations extend the remote-sensing picture of Jupiter's
ring system into regions where the dust is too tenuous to be readily detected
optically. Juno Waves reveals that the innermost dust environment contains
fine radial, vertical, and hemispheric structure superposed on the
large-scale Jovian halo, while simultaneously providing a statistical view
of nonlinear plasma structures throughout Jupiter's polar magnetosphere.

\appendix

\section{Representative Waveforms of Classified Transient Signals}
\label{app:waveform_examples}

\begin{figure*}[htb!]
    \centering
    \includegraphics[width=\textwidth]
    {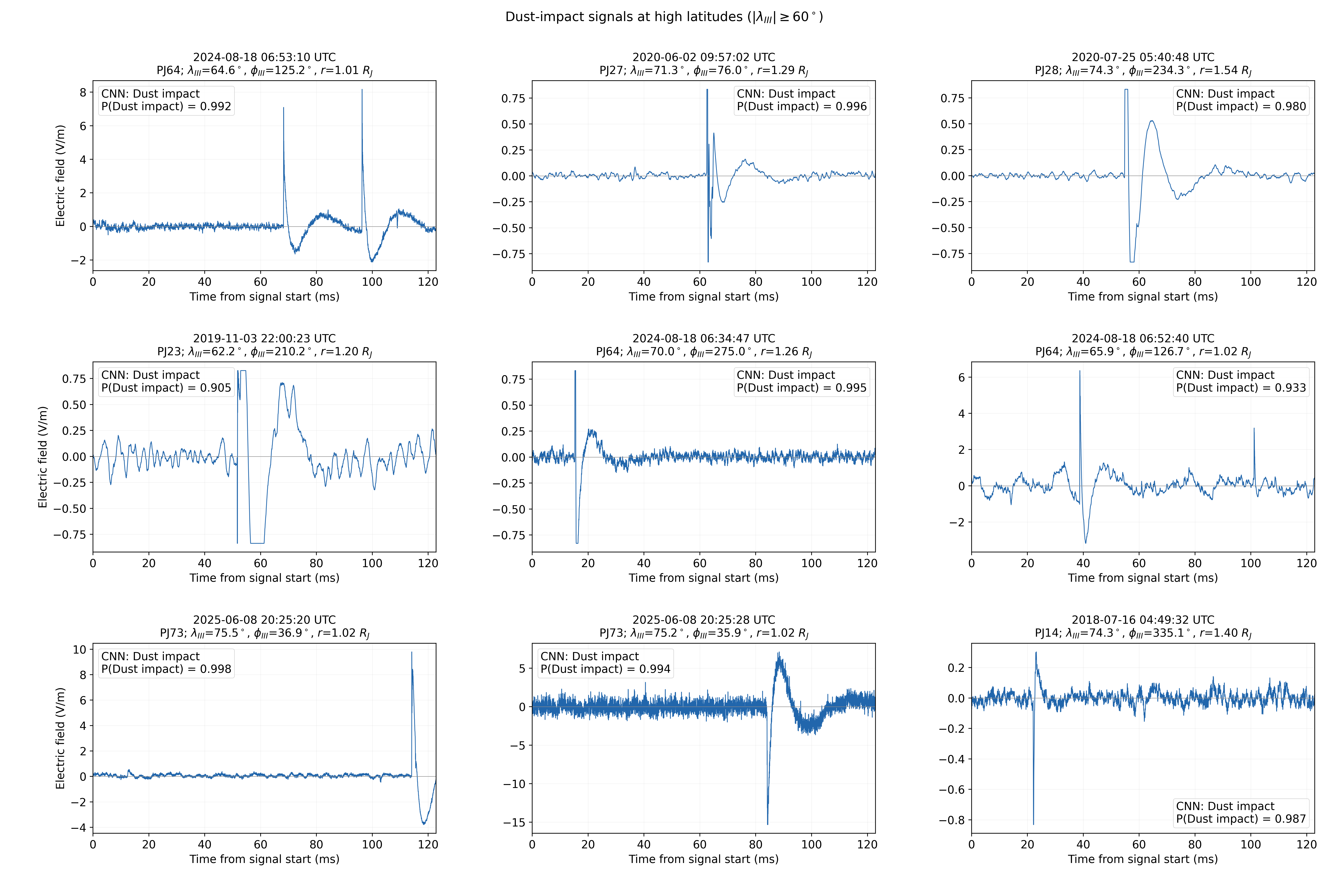}
    \caption{
    Representative dust-impact signals identified at high magnetic latitudes
    ($|\lambda_{\rm III}|\geq60^\circ$). Each panel shows the electric-field
    waveform recorded by Juno Waves together with the observation time,
    perijove number, magnetic latitude $\lambda_{\rm III}$, System III
    longitude $\phi_{\rm III}$, radial distance $r$, and CNN classification
    probability. The selected examples illustrate the impulsive signatures
    classified as dust impacts within the otherwise sparse high-latitude
    dust population.
    }
    \label{fig:appendix_dust_high_lat}
\end{figure*}

\begin{figure*}[htb!]
    \centering
    \includegraphics[width=\textwidth]
    {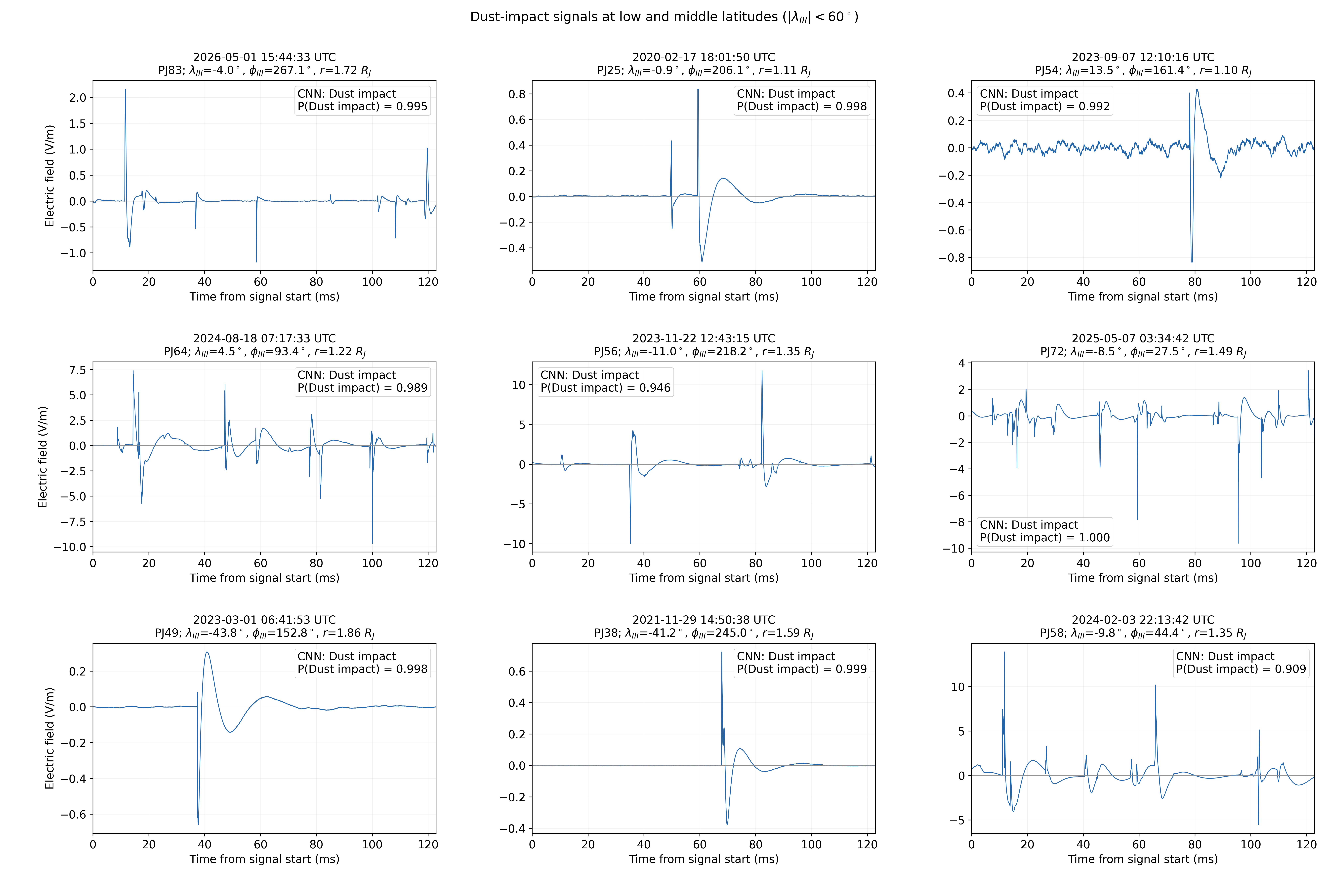}
    \caption{
    Representative dust-impact signals identified at low and middle magnetic
    latitudes ($|\lambda_{\rm III}|<60^\circ$). The examples demonstrate the
    range of waveform morphologies and amplitudes present within the
    dust-impact class. Quantities listed above each waveform are defined as
    in Figure~\ref{fig:appendix_dust_high_lat}.
    }
    \label{fig:appendix_dust_low_mid_lat}
\end{figure*}

\begin{figure*}[htb!]
    \centering
    \includegraphics[width=\textwidth]
    {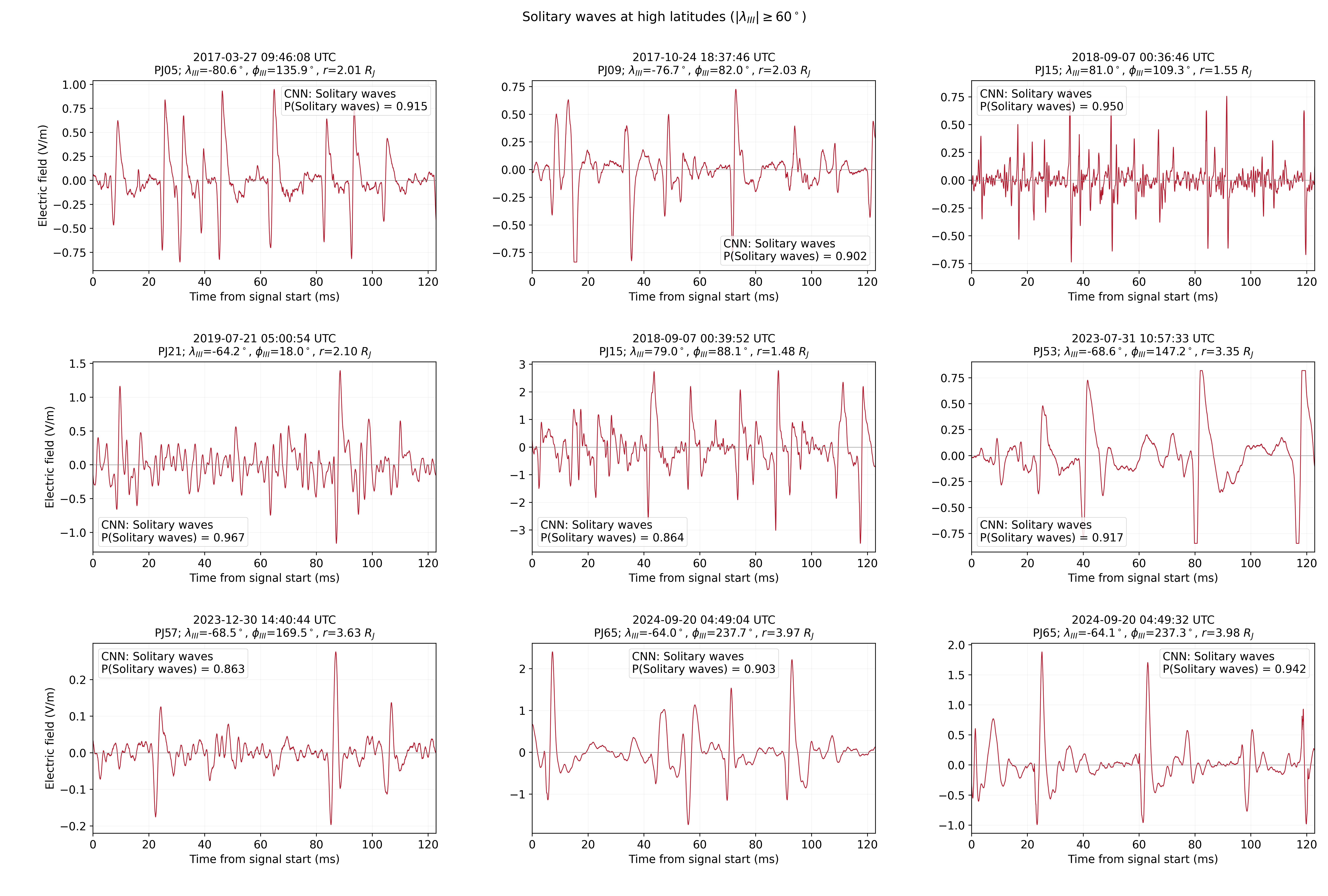}
    \caption{
    Representative high-latitude solitary-wave signals identified by the
    multi-class CNN. The selected waveforms illustrate repeated localized
    electric-field structures characteristic of the solitary-wave class.
    Quantities listed above each waveform are defined as in
    Figure~\ref{fig:appendix_dust_high_lat}.
    }
    \label{fig:appendix_solitary}
\end{figure*}

\begin{figure*}[htb!]
    \centering
    \includegraphics[width=\textwidth]
    {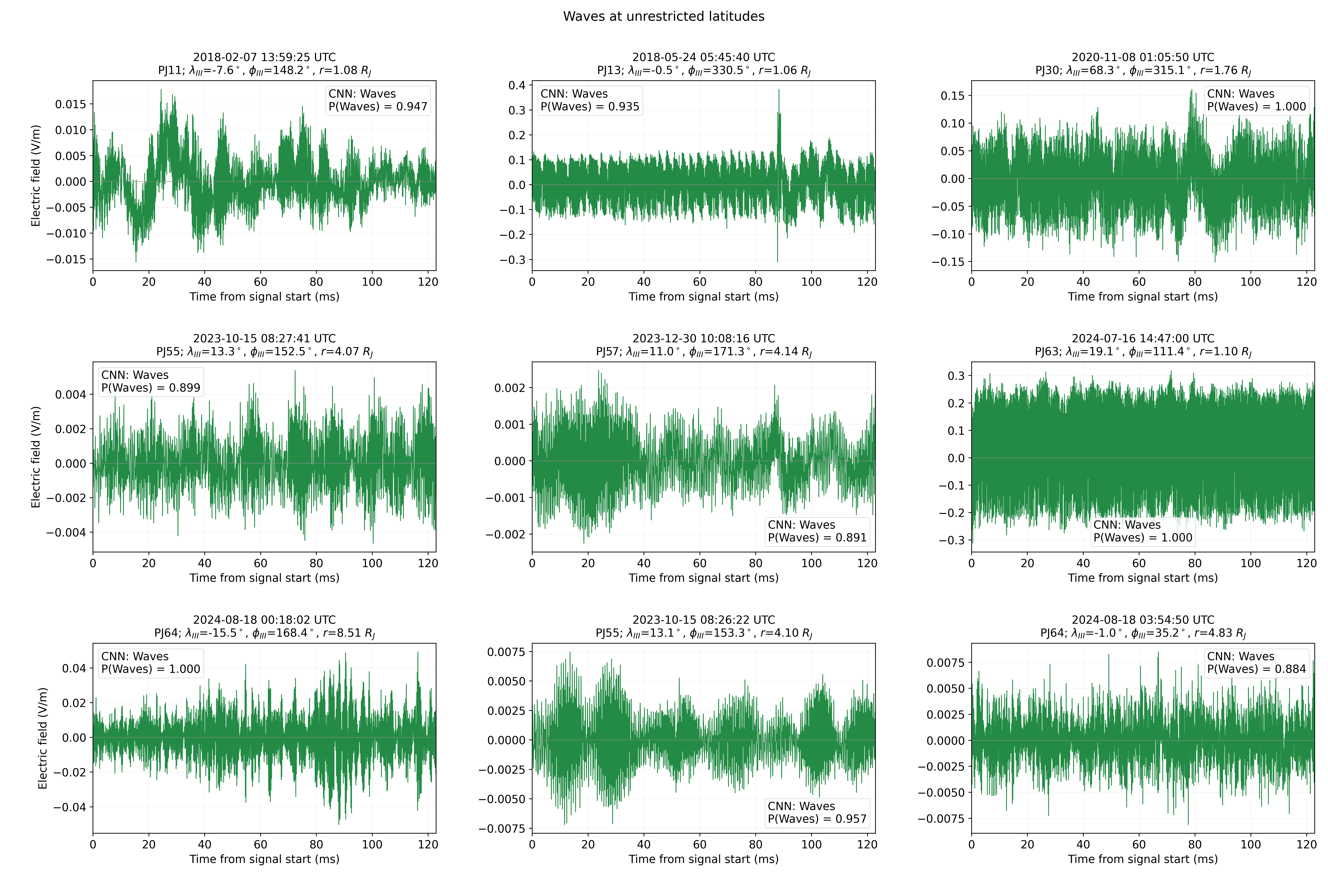}
    \caption{
    Representative plasma-wave signals identified by the multi-class CNN.
    These examples span a broad range of magnetic latitudes and radial
    distances and illustrate the more continuous oscillatory electric-field
    signatures assigned to the other plasma-wave class. Quantities listed
    above each waveform are defined as in
    Figure~\ref{fig:appendix_dust_high_lat}.
    }
    \label{fig:appendix_waves}
\end{figure*}

To provide representative examples of the transient signals identified by the multi-class CNN, Figures~\ref{fig:appendix_dust_high_lat}--\ref{fig:appendix_waves} show selected Juno Waves electric-field waveforms classified as dust impacts, solitary waves, and other plasma waves. Each panel lists the observation time, perijove number, spacecraft magnetic latitude and System III longitude, radial distance from Jupiter, and the classification probability returned by the CNN.

Figure~\ref{fig:appendix_dust_high_lat} shows representative dust-impact signals detected at high magnetic latitudes ($|\lambda_{\rm III}|\geq60^\circ$). Despite their locations well outside the dominant low-latitude dust population described in Section~\ref{sec:dust_global_distribution}, these events exhibit impulsive electric-field signatures characteristic of the dust-impact class and are identified by the CNN with high confidence. The examples span both hemispheres, multiple perijoves, and radial distances from approximately $1.0$ to $1.5~R_J$, demonstrating that the small high-latitude dust population in the global distributions is not restricted to a single orbit or observing interval.

For comparison, Figure~\ref{fig:appendix_dust_low_mid_lat} presents dust-impact signals detected at low and middle magnetic latitudes ($|\lambda_{\rm III}|<60^\circ$), where the majority of the dust population is observed. The waveforms span a range of amplitudes and morphologies, including isolated impulsive signatures as well as observations containing multiple sharp impulses and subsequent relaxation structures. The CNN assigns high dust-impact probabilities to these examples despite this substantial waveform-to-waveform variability.

Figure~\ref{fig:appendix_solitary} shows representative solitary waves detected at high magnetic latitudes. In contrast to the dust-impact examples, these signals commonly contain sequences of repeated bipolar or impulsive electric-field structures distributed throughout the waveform interval. The examples are drawn from multiple perijoves and cover radial distances from approximately $1.5$ to $4.0~R_J$. Their predominantly high-latitude locations are consistent with the polar concentration of solitary waves shown in Section~\ref{sec:solitary_distribution}.

Finally, Figure~\ref{fig:appendix_waves} shows representative examples assigned to the other plasma-wave class. These signals are generally characterized by more continuous oscillatory activity over a substantial fraction of the waveform interval, distinguishing them from the isolated impulsive structures typical of dust impacts and the localized structures associated with the solitary-wave class. The examples span a broad range of magnetic latitudes and radial distances, consistent with the more spatially distributed occurrence of this signal population.

\section{Charged-Dust Timescales and Magnetic-Equator Geometry}
\label{app:dust_dynamics_support}

To provide additional context for the interpretation of the vertical dust distribution, we performed two complementary calculations. The first compares the dynamical lifetime of charged grains with their characteristic charging time, while the second evaluates the radial variation of Jupiter's magnetic equator relative to the System~III equatorial plane. These calculations are intended to characterize relevant physical timescales and geometrical trends rather than to provide a complete three-dimensional model of the observed dust distribution.

\subsection{Dust Lifetime and Charging Time}

We first estimated the characteristic lifetime of dust grains undergoing electromagnetic transport in Jupiter's inner magnetosphere. Test particles were simulated for grain radii between $0.1$ and $5~\mu$m. To improve the statistics of the rapidly evolving small-grain population, 5000 independent particles were simulated at each grain size for $a\leq0.5~\mu$m, while 30 particles were used at each larger grain size. Initial System~III longitudes were uniformly distributed between $0^\circ$ and $360^\circ$, and initial radial distances were randomly selected between $1.71$ and $1.85~R_J$. The grains were assumed to be spherical with density $\rho_{\rm d}=3000~{\rm kg\,m^{-3}}$, and each particle was initialized on a prograde circular orbit with the local Keplerian velocity and with zero initial charge, $Q_0=0$.

The trajectory calculation includes Jupiter's gravitational field, the JRM33 internal magnetic field, the Con2020 external magnetodisc field, the corotation electric field, and time-dependent grain charging. Each trajectory was followed until the grain reached $r=1.01~R_J$, which was taken to represent removal by impact with Jupiter. The elapsed time to this boundary defines the grain lifetime, $T$. The mean and standard deviation were then calculated from the simulated trajectory ensemble at each grain radius.

Over the modeled grain-size range, the simulated lifetimes are approximately described by
\begin{equation}
T \approx 71.1\,a^{3.23}\ {\rm days},
\label{eq:dust_lifetime_fit}
\end{equation}
where $a$ is expressed in $\mu$m. This dependence is broadly consistent with the approximate relation
\begin{equation}
T \sim 100\,a^3\ {\rm days}
\end{equation}
reported by \citet{horanyi_plasma_2010}. The agreement is primarily in the strong grain-size dependence rather than in the exact normalization.

A separate calculation was performed to estimate the charging timescale. For the same grain-size and initial radial ranges, particles were initialized with $Q(0)=0$, but their spatial positions were held fixed in order to
isolate the local charging process from orbital evolution. The grain charge was evolved under the adopted photoelectron-emission and ambient-electron collection model until it first reached 95\% of the local equilibrium value, $Q_{\rm eq}$. The corresponding time is defined as $t_{\rm charge}$.

\begin{figure*}[htb!]
    \centering
    \includegraphics[width=0.7\textwidth]
    {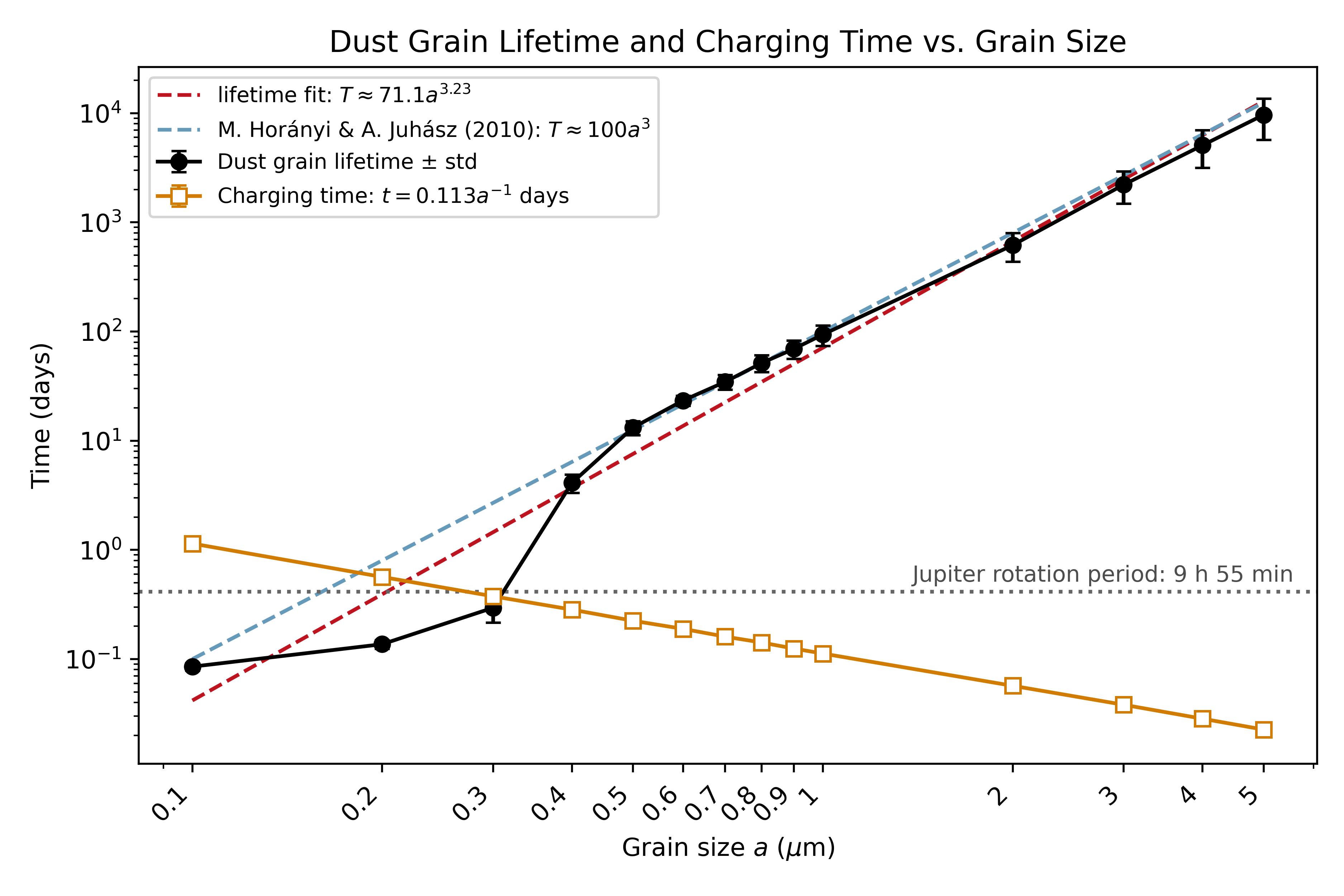}

    \caption{
    Simulated dust-grain lifetime and charging time as functions of grain
    radius. Black circles show the mean dynamical lifetime, with error bars
    indicating the standard deviation. For $a\leq0.5~\mu$m, 5000 test
    particles were simulated at each grain size; 30 particles were used at
    each larger grain size. The red dashed curve shows the empirical fit
    $T=71.1a^{3.23}$ days, while the blue dashed curve shows the approximate
    $T=100a^3$ day relation from \citet{horanyi_plasma_2010}. Orange squares
    show the time required for a grain initially at $Q=0$ to reach 95\% of
    the local equilibrium charge, approximated by
    $t_{\rm charge}=0.113a^{-1}$ days. The horizontal dotted line indicates
    Jupiter's System~III rotation period ($\sim9$~h~55~min).
    }
    \label{fig:dust_lifetime_charging}
\end{figure*}

The resulting charging times are approximately described by
\begin{equation}
t_{\rm charge}
\approx
0.113\,a^{-1}\ {\rm days},
\label{eq:dust_charging_time}
\end{equation}
again with $a$ in $\mu$m. Figure~\ref{fig:dust_lifetime_charging} compares the two timescales together with Jupiter's System~III rotation period of approximately 9~h~55~min.

The comparison reveals two qualitatively different dynamical regimes. For grains with $a\lesssim0.3~\mu$m, the simulated mean lifetime is shorter than Jupiter's System~III rotation period of approximately 9~h~55~min. These particles are therefore typically removed before experiencing one complete rotation of Jupiter's non-axisymmetric magnetic field. Although they remain subject to instantaneous Lorentz forces, their short lifetimes make sustained phase-coherent interaction with the rotating field, including the development of Lorentz-resonant orbital evolution, unlikely.

The same small grains also have comparatively long charging times. For $a\sim0.1$--$0.3~\mu$m, the time required to reach 95\% of the local equilibrium charge is comparable to or longer than the simulated lifetime. They may therefore be lost before either reaching local charging equilibrium or sampling a complete Jovian rotation. Their trajectories represent a rapid-loss, nonequilibrium-charging regime in which the instantaneous electromagnetic force can be strong, owing to their large charge-to-mass ratios, but the assumptions underlying long-term Lorentz-resonance transport are not necessarily satisfied.

A sharp change occurs for grains with $a\gtrsim0.4~\mu$m. Their simulated lifetimes exceed the Jovian rotation period by more than an order of magnitude, while their charging times become much shorter than their lifetimes. These grains can therefore experience many rotations of the magnetic-field pattern after approaching local charging equilibrium, making long-term electromagnetic transport and Lorentz-resonant effects much more dynamically relevant.

The lifetime relation should be regarded as preliminary. A subset of the simulated trajectories, particularly at larger grain sizes, did not reach $r=1.01~R_J$ before the adopted maximum integration time and are therefore right-censored. The mean lifetimes and the power-law fit in Figure~\ref{fig:dust_lifetime_charging} do not explicitly account for this censoring. A more complete treatment would require survival-analysis methods or substantially longer integrations. The present calculation is therefore used only to establish the approximate ordering of the dynamical and charging timescales.

\subsection{Mean Magnetic Equator Displacement}

Because electromagnetic forces play an important role in the dynamics of small grains, we also examined the location of Jupiter's magnetic equator relative to the System~III equatorial plane. The magnetic field was calculated using the JRM33 internal field model together with the Con2020 external magnetodisc field model. We evaluated the magnetic equator using two definitions, corresponding to $B_r=0$ and $B_\rho=0$, respectively. For each definition, the magnetic-equator position was evaluated over 360 equally spaced System~III longitudes and averaged to obtain the mean vertical displacement, $Z_{\rm ME}$, as a function of System~III cylindrical distance.

\begin{figure*}[htb!]
    \centering
    \includegraphics[width=0.7\textwidth]
    {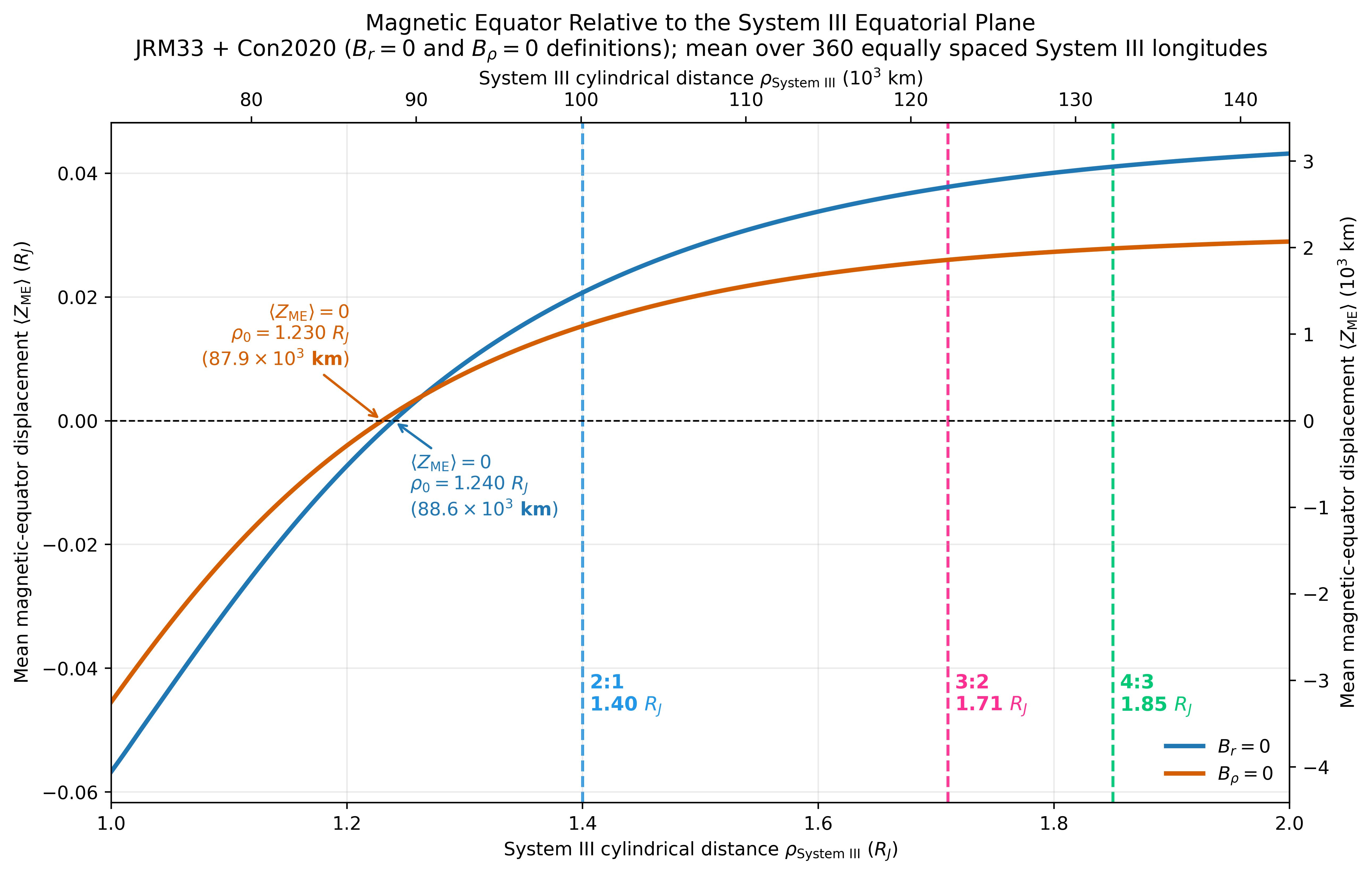}

    \caption{
    Longitude-averaged displacement of Jupiter's magnetic equator relative
    to the System~III equatorial plane as a function of System~III
    cylindrical distance. The magnetic field is calculated using JRM33
    together with Con2020 and averaged over 360 equally spaced System~III
    longitudes. Blue and orange curves show magnetic equator locations
    defined by $B_r=0$ and $B_\rho=0$, respectively. Both definitions show
    the same radial trend, crossing the System~III equatorial plane near
    $\rho=1.23$--$1.24~R_J$ and shifting progressively northward at larger
    radial distances. Vertical dashed lines mark the approximate locations
    of the $2{:}1$, $3{:}2$, and $4{:}3$ Lorentz resonances.
    }
    \label{fig:magnetic_equator_offset}
\end{figure*}

Figure~\ref{fig:magnetic_equator_offset} shows that the two definitions produce very similar magnetic equator geometries and the same overall radial trend. At the smallest cylindrical distances, the mean magnetic equator lies south of the System~III equatorial plane. It then crosses to the north near $\rho\sim1.23$--$1.24~R_J$, with zero crossings at approximately $1.240~R_J$ for the $B_r=0$ definition and $1.230~R_J$ for the $B_\rho=0$ definition. At larger radial distances, both curves shift progressively northward, reaching displacements of order $0.02~R_J$ near the $2{:}1$ Lorentz resonance and several $10^{-2}~R_J$ toward the main-ring region. The close agreement between the two definitions indicates that the inferred radial trend is not sensitive to the particular magnetic equator definition adopted.

The radial evolution of $Z_{\rm ME}$ is notable because both magnetic-equator definitions place the transition from southward to northward displacement near the radial range where the observed dust distribution changes from an approximately symmetric inner layer to a northward-broadened population, $\rho\sim1.2$--$1.3~R_J$. The magnetic-equator displacement is much smaller than the total thickness of the observed halo; for example, the displacement near $1.4~R_J$ is only of order $\sim0.02~R_J$, compared with an observed halo HWHM of approximately $0.367~R_J$. The magnetic equator therefore cannot by itself account for the large vertical extent of the dust population. Instead, its radial displacement provides a symmetry-breaking reference geometry that may bias the vertical Lorentz forcing of charged grains toward one hemisphere.

Taken together, the charging-timescale and magnetic-equator calculations illustrate two complications that are not captured by a simple equilibrium-charge, axisymmetric dust model. The smallest grains may undergo substantial orbital evolution while remaining out of local charging equilibrium, while the magnetic reference surface itself varies relative to the System~III equatorial plane with radial distance. Both effects can contribute to radius-dependent vertical structure, although a quantitative comparison with the Juno Waves observations will require full three-dimensional dust-dynamics simulations with time-dependent charging and the actual spacecraft sampling geometry.

\section{Perijove-Dependent Effective Impact Area and Spacecraft Speed}
\label{app:impact_area_speed}

\begin{figure*}[htb!]
    \centering
    \includegraphics[width=0.7\textwidth]
    {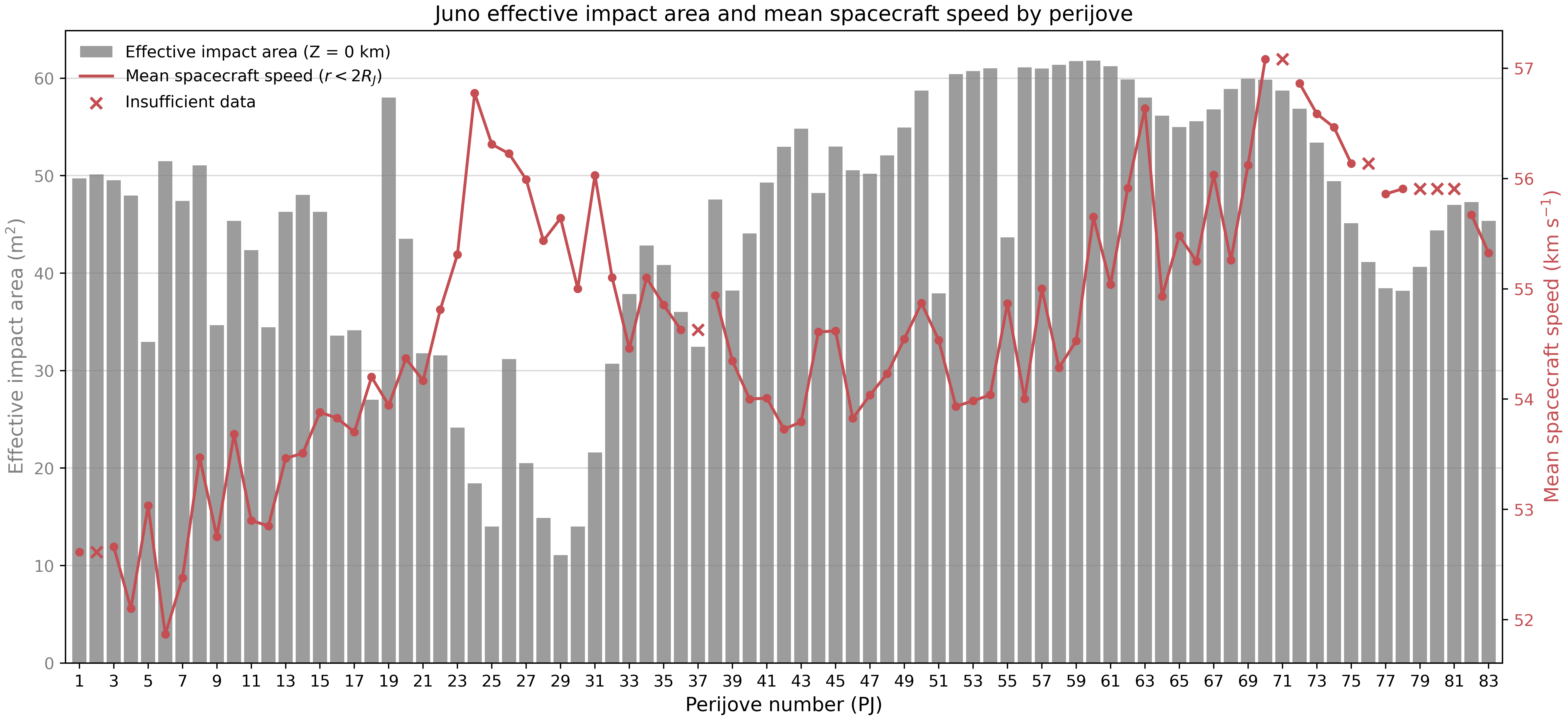}

    \caption{
    Effective impact area and mean Juno spacecraft speed as a function of
    perijove number. Gray bars show the effective impact area evaluated for
    each perijove at $Z=0$, and red circles show the mean spacecraft speed
    for observations within $r<2~R_J$. Red crosses mark PJ2, PJ37, PJ71,
    PJ76, PJ79, PJ80, and PJ81, for which insufficient near-perijove data
    are available to determine a representative mean speed.
    }
    \label{fig:pj_area_speed}
\end{figure*}

The effective dust impact area was determined using a high-fidelity three-dimensional model of the Juno spacecraft. Because the spacecraft is approximately symmetric about its spin axis (the spacecraft +Z axis), its projected cross-sectional area can be parameterized as a function of the angle between the incoming dust direction and this axis. Reference cross-sectional areas were calculated from the spacecraft model at uniformly sampled angles relative to the +Z axis and in multiple sampling planes parallel to this axis. A parametric fit reproduces these reference cross-sectional areas to within 2.96\%. The resulting model was used to determine the effective dust impact area for each perijove.

Figure~\ref{fig:pj_area_speed} summarizes the effective impact area adopted
for each perijove together with the mean Juno spacecraft speed within
$r<2~R_J$. The effective impact area varies substantially among perijoves,
from approximately $10$ to more than $60~{\rm m^2}$, reflecting changes in
the spacecraft geometry relative to the incoming dust population. The mean
spacecraft speed within $2~R_J$ is much less variable, typically ranging
from approximately $52$ to $57~{\rm km\,s^{-1}}$.

For PJ2, PJ37, PJ71, PJ76, PJ79, PJ80, and PJ81, insufficient near-perijove
data are available, these cases are marked by crosses in
Figure~\ref{fig:pj_area_speed}. The perijove-dependent effective areas are
used in the number-density calculation described in
Section~\ref{sec:dust_global_distribution}.

\begin{acknowledgments}
This research was supported by the W. M. Keck Foundation and the NASA Juno mission.
\end{acknowledgments}

\bibliographystyle{aasjournalv7}
\bibliography{references}{}



\end{document}